\documentclass{aa}
\usepackage{graphicx}
\usepackage{natbib}
\begin{document}
\title{The intrinsic emission of Seyfert galaxies observed
 with BeppoSAX/PDS}
\subtitle{I. Comparison of the average spectra of the three classes of Seyfert Galaxies}
\author{S. Deluit
\inst{1,2}
\and T. Courvoisier\inst{1,2}}
\institute{\textit{INTEGRAL} Science Data Center, 16 Chemin d'Ecogia, ch-1290 Versoix, Switzerland
\and Geneva Observatory, 51 Chemin des Maillettes, ch-1290 Sauverny, Switzerland}
\offprints{S. Deluit,\\ \email{Sandrine.Deluit@obs.unige.ch}}
\date{Received/Accepted}
\abstract{
We present a  study of the hard X-ray spectrum ($>$15 keV) of  different classes of Seyfert galaxies observed
 with BeppoSAX/PDS. 
Using hard X-ray data, we avoid absorption effects modifying the Seyfert emission and  have direct access to
 the central
engine  of these sources.
The aim of this study is first to characterize the general properties of the hard X-ray
spectrum of Seyfert 1, 1.5 and 2 galaxies and secondly to compare their intrinsic emission  
to test unified models according to which all the classes have the same nucleus.\\
We compute the average spectrum of 14 Sy 1, 9 Sy 1.5 and 22 Sy 2 galaxies  observed by the PDS (15-136 keV). 
The average spectrum of Sy 1 differs from that of Sy 2, the first requiring the
presence of a high energy cutoff which  is absent in the second. We also show that
the reflection component is possibly more important in the Sy 2 emission.
The nature of Sy 1.5 galaxies is ambiguous as they have a
negligible reflection component (like Sy 1) and  do not require a cutoff (like Sy 2).
}
\titlerunning{The intrinsic emission of Seyfert galaxies}
\authorrunning{Sandrine Deluit \& Thierry Courvoisier}
\maketitle
\section{Introduction}
According  to unified models \citep{Antonucci}, Seyfert 2 galaxies harbor a bright Seyfert 1 nucleus which is hidden from our view by an optically and geometrically thick obscuring torus.
Consequently, the difference between type 1 and 2 Seyfert galaxies would be only due to viewing  angle. 
Several observational facts in favor of unified models have been found:
\begin{itemize}
\item detection of Polarized Broad Lines (PBL hereafter) in several Seyfert 2 galaxies
\citep{Antonucci,MillerGood,Tran95,Young,Awaki,Moran2000,Alexander,GuPbl,Lumsden}, interpreted as
scattering of the Broad Lines Region emission by warm material placed above an absorption torus. 
\item large absorbing column densities observed in the hard X-ray spectra of Seyfert 2 galaxies
(\citet{Guainazzi2001} and references therein).
\item suspicion of anisotropic ionization radiation from the nucleus through the structures in the
light of [OIII] line at 5007 $\textrm{\AA}$ (e.g. in NGC 5252 \citep{Tadhunter}).
\end{itemize}
 X-ray data have played a 
major role in support of  unified models. 
The X-ray spectra of Seyfert galaxies are known relatively well and are composed at energies above 3 keV
of a  power law, a Fe line and a Compton reflection component (\citet{NP}, \citet{Gondek}
\citet{Nandra97}, \citet{Z99}, \citet{Z2000}). Some questions remain  like the shape and the presence
of a cutoff at higher energies. 
Furthermore, some observations are inconsistent with unified models at least in their simplest formulation. High energy
 spectra of a sample of Seyfert 2 galaxies observed by OSSE suggest an intrinsic difference in the
 hard X-ray emission of both classes of Seyfert galaxies. The two classes differ in the slope  of the 
 intrinsic
 power law, in the energy of the exponential cutoff, in the amount of absorption and reflection 
 (Zdziarski, 1995).\\
 Hard X-ray spectra are a powerful tool to probe 
and test  unification models by providing a direct view on the central engine emission mostly free of absorption effects. 
The form of the individual hard X-ray spectra is poorly  constrained due to limited
photon statistics. This can be improved by considering the average spectrum of a sample to extract the
 properties of a class.\\
We study the hard X-ray emission (above 15 keV) of Seyfert 1, 1.5 and 2 galaxies in order to investigate
several questions:
\begin{itemize}
\item Are the high energy index slope of Seyfert 2 and  Seyfert 1 identical? 
\item Do the Seyfert 1 and 2 spectra have a cutoff? In thermal Compton models
 the high energy cutoff provides a reliable estimate of the comptonizing  material temperature \citep{Rybicki}. 
 Therefore, differences in  cutoffs or their absence indicate
  different comptonizing medium temperatures.
\item What is the importance of Compton reprocessed emission for each class? Could the
reflection  be the only cause to the observed difference between Seyfert 1 and Seyfert 2?
\end{itemize} 
 To answer these questions, we compute the average PDS spectrum of each class. We consider for inclusion in the Seyfert 2 sample only galaxies for which the
   absorption 
  column density is less than 7$\cdot$10$^{23}$cm$^{-2}$ to ensure that, above 15 keV, absorption  is 
   less than 50$\%$ of the flux.  
   We are thus in the position to characterize the intrinsic emission of the various types of objects.
\section{ Selection procedure of our sample }
We start by including all (i.e. 68) Seyfert PDS spectra available in the public archive of BeppoSAX Data
Center\footnote{http://www.asdc.asi.it/bepposax/}.  We have investigated the possibility of source confusion 
in the PDS aperture by looking in the MECS images and excluded two confused objects from this study. 
We then form three subsamples characterizing Sy 1, Sy 1.5 and Sy 2 galaxies respectively. 
\subsection{The nature of  objects composing the subsamples}
We follow the optical classification of Seyfert galaxies indicated in the NASA/IPAC
Extragalactic Database\footnote{http://nedwww.ipac.caltech.edu/}(NED) and  exclude 
LINERs for the ambiguity of their classification. We include Sy 
1 and Sy 1.2 galaxies in the Seyfert 1 subsample, Sy 1.5 in the Seyfert 1.5 subsample and
Sy 1.9 and Sy 2  in  the Seyfert 2 subsample. We exclude Seyfert 1.8 for
which the classification is too ambiguous between Seyfert 1.5 and Seyfert 2 galaxies. 
\subsection{Absorption effects in Seyfert 2 galaxies emission}
Hard X-rays give a direct access to the intrinsic emission
provided that we  avoid absorption effects which  contaminate the Seyfert 2 galaxies spectra  at
 lower energies.  
 Objects with  N$_{\footnotesize{\textrm{H}}}$$>$1.5$\cdot$10$^{24}$cm$^{-2}$ (so called Compton thick) have
  a Thomson
  optical depth larger than one. Electron scattering modifies their emission even in the hard X-ray energy domain. We
   therefore eliminated
   all such
   objects from consideration. 
In summary, we consider only objects for which   e$^{-\sigma_{T}N_{H}} \times
e^{-\sigma_{ph}(E) N_{H}}$ is larger than 0.5 at 15 keV, $\sigma_{T}$ and $\sigma_{ph}$
are respectively the Thomson and the photoelectric cross section. This means that we keep only objects
having a column density  less than 7$\cdot$10$^{23}$cm$^{-2}$. \\
The role of absorption in our Seyfert 2 class study is  exhaustively explained in the appendix B.\\
 We present in Figure 1  the column density distribution for the  Seyfert 2
galaxies sample. The majority of the objects have N$_{\footnotesize{\textrm{H}}}$ between 10$^{22}$cm$^{-2}$ and 
 10$^{23}$cm$^{-2}$.

\begin{figure}[!h]
\centering
\includegraphics[width=8.2cm]{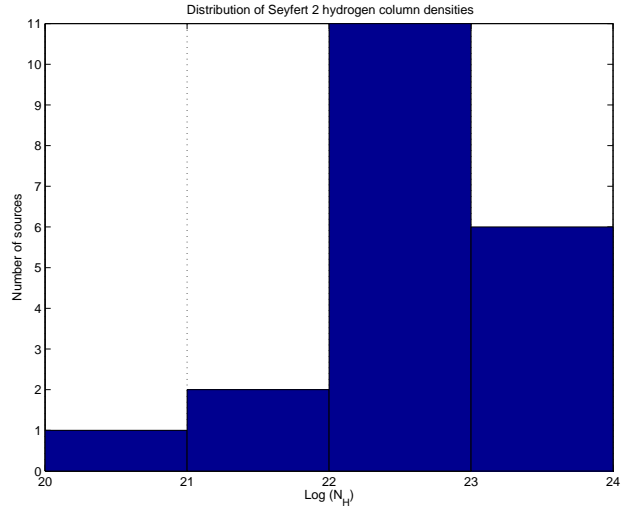}
\caption{Distribution of the hydrogen column density of sources composing
our Seyfert 2 sample}
\end{figure}

We  present in  Figure 2 the hydrogen column density distribution versus the 
redshift for  Seyfert 2 galaxies.
\begin{figure}[!h]
\centering
\includegraphics[width=8cm]{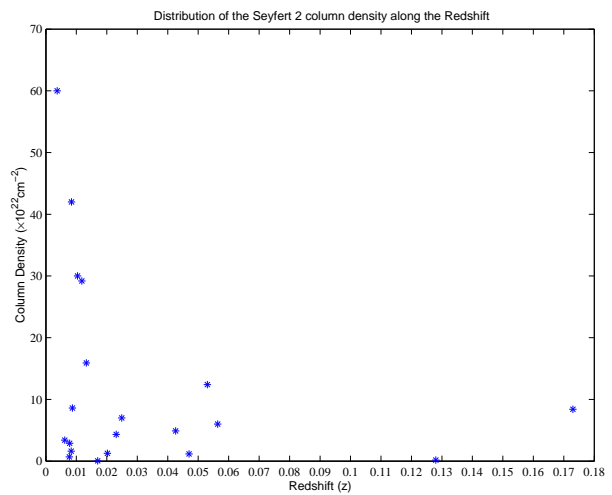} 
\caption{Distribution of Seyfert 2 hydrogen column density along the redshift}
\end{figure}\\
We note a large number of high column densities at low redshift. This is probably an observational artefact due to
 the fact
that distant and highly absorbed Seyfert 2 galaxies are hardly detectable.
\subsection{Data quality criterion}
 The third criterion we introduce is the quality of the data. We calculate the \textit{integrated}
 signal to noise ratio  of each object in the
PDS energy range (15-136 keV). The data are used only when the integrated
signal to noise ratio is equal to or larger than 2$\sigma$. This threshold is high enough  to conclude to a  
detection by the PDS detector for a known source.
\section{Properties of our samples}  
\subsection{Redshift distributions }
We present in Figure 3 the redshift distribution for the three classes of
 Seyfert galaxies.

\begin{figure}[!h]
\centering
\includegraphics[width=8.5cm, height=8cm]{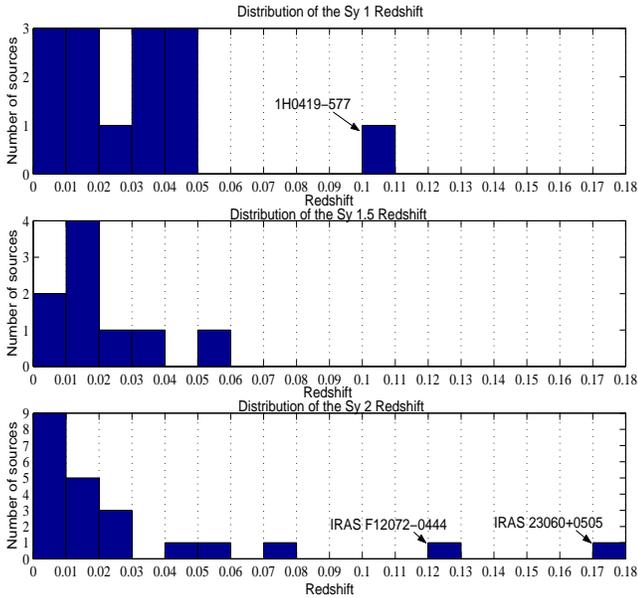} 
\caption{Redshift distribution for the three classes of Seyfert galaxies}
\end{figure}

Most of the objects have a redshift less than 0.05 (93$\%$ for Sy 1, 89$\%$ for Sy 1.5 and 82$\%$ for Sy 2 galaxies). The difference between the redshift distributions
 of Sy 1 and Sy 2 galaxies may  be due to biases in object selections when planning
X-ray observation programmes. Indeed, absorbed objects being more difficult to observe will be looked for at small 
distances. 
\subsection{Comparison of the data quality }
We determine the integrated signal to noise ratio in the PDS energy range (15-136 keV) for each object (Figure 4).
\begin{figure}[!h]
\centering
\includegraphics[width=8.5cm, height=8cm]{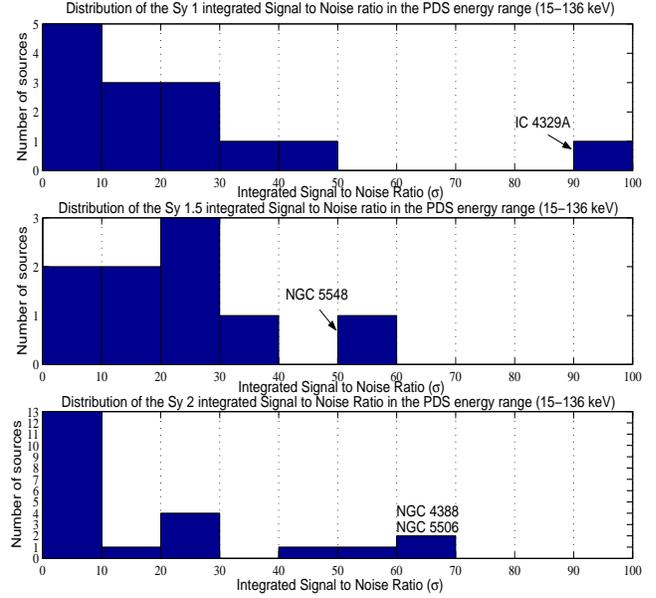}
\caption{Distribution of the integrated signal to noise ratio  for the three classes of 
Seyfert in the PDS energy range}
\end{figure}

The majority of the objects have an integrated signal to noise
ratio less than 30$\sigma$ (79$\%$ for Sy 1, 78$\%$ for Sy 1.5 and 82$\%$ for Sy 2 galaxies). 
The Seyfert 1, 1.5 and 2 samples are dominated by low signal to noise spectra.
\subsection{Flux distributions} 
We measure the integrated flux in the PDS energy range of each object (Figure 5).
The majority of the sources have a hard X-ray flux less than 
15$\cdot$10$^{-11}$erg\,s$^{-1}$\,cm$^{-2}$ between
 15 keV and 136
keV. 
\begin{figure}[!h]
\centering
\includegraphics[width=8.5cm, height=8.2cm]{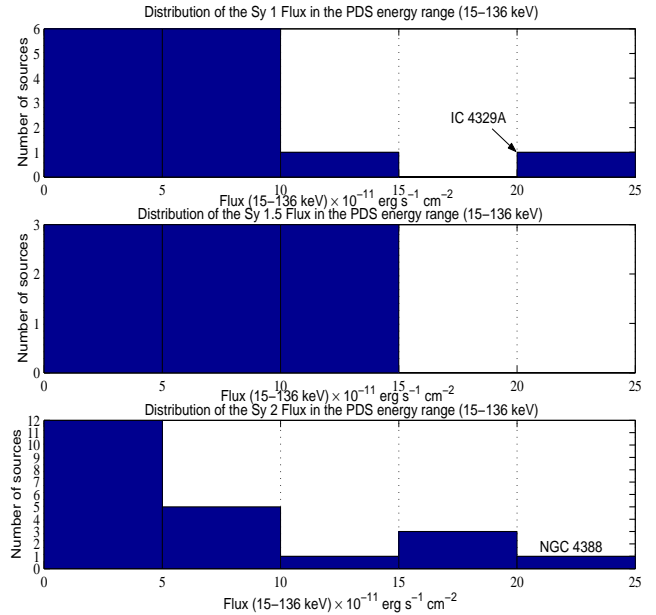}
\caption{Flux distribution for the three classes of Seyfert in the PDS energy range}
\end{figure}

The proportion of objects with a flux less than or equal to
15$\cdot$10$^{-11}$erg\,s$^{-1}$\,cm$^{-2}$ is of 93$\%$ for Sy 1, 100$\%$ for Sy 1.5 and of 82$\%$
for Sy 2 galaxies.
\subsection{ Luminosity distributions} 
From the flux  we calculate the integrated luminosity using H$_{0}$=75
km\,s$^{-1}$\,Mpc$^{-1}$ and q$_{0}$=0.5 as cosmological
 parameters. The results are shown in Figure 6.
\begin{figure}[!h]
\centering
\includegraphics[width=8.8cm, height=8.5cm]{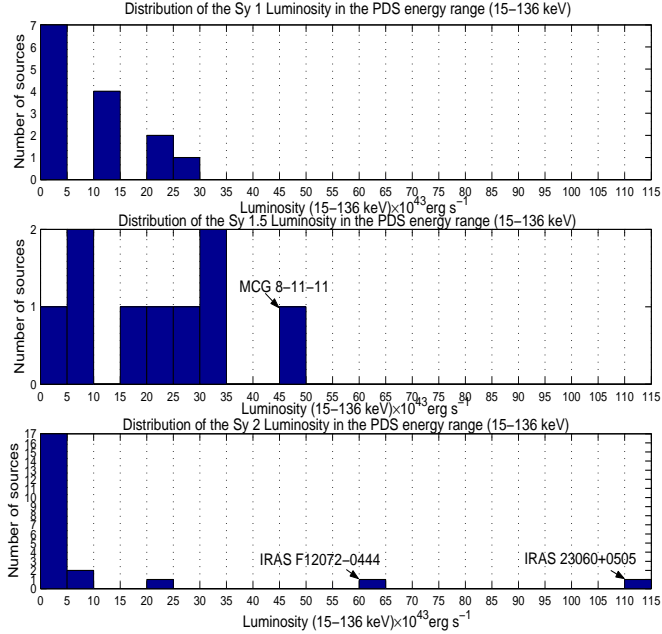}
\caption{Luminosity distribution for the three classes of Seyfert in the PDS energy range}
\end{figure}

The majority of the objects have a hard X-ray luminosity less than  or equal
to 30$\cdot$10$^{43}$erg\,s$^{-1}$
 between 15 keV and 136
keV (93$\%$ for Sy 1, 100$\%$ for Sy 1.5 and 91$\%$ for Sy 2). 
In the Seyfert 2 subsample, two objects have a
luminosity  above 30$\cdot$10$^{43}$erg\,s$^{-1}$: IRAS F12072-0444 with a luminosity of
61.5$\cdot$10$^{43}$erg\,s$^{-1}$ and  IRAS 23060+0505,  with  a luminosity of 
111$\cdot$10$^{43}$erg\,s$^{-1}$. These two galaxies are warm Ultra Luminous Infrared Galaxies and are probably
 quasars regarding their hard X-ray luminosity. However, they are classified as Seyfert
  2 galaxies in NED and therefore kept in our sample.
\subsection{Excluded objects}
Several objects have been excluded from our initial sample. We review below all the dismissed sources and the
 reason of their exclusion.
\begin{enumerate}
\item \textbf{NGC 4151:} with an integrated signal to noise ratio of 227.5$\sigma$ and an integrated flux in the
 PDS energy range of
58$\cdot$10$^{-11}$erg\,s$^{-1}$\,cm$^{-2}$, this object would strongly dominate the sample. We  will discuss this
  bright Seyfert 1.5
galaxy in the paper II of this study where individual objects are discussed \citep{Deluit2}.
\item \textbf{Mrk 1073:} this object is not detected in the MECS and LECS instruments and has a
very high signal in the PDS. We excluded it from our sample because its PDS spectrum
is probably in part due to a nearby source. 
\item \textbf{NGC 1386}:  a large fraction of the PDS flux is due to the nearby NGC 1365 \citep{Maiolino}. 
\item \textbf{MCG 5-18-002}: we excluded this source in reason  of its  
uncertain absorption given as N$_{\footnotesize{\textrm{H}}}$ $>$10$^{25}$\,cm$^{-2}$ by
\citet{Maiolino}. The value is based only on the [OIII] flux and requires further confirmation.
\item Considering the data quality  criterion, we also exclude four Seyfert 1 (\textbf{TON S180, GQCOMAE,
IRAS 13224-3809 and Mrk 478}) and four Seyfert 2 (\textbf{NGC 3081, IRAS 11058-1131, IRAS 10126+7339 and
NGC 4941}). All these objects have an integrated signal to noise ratio less than 2$\sigma$.
\end{enumerate}
\subsection{The final sample}
The initial sample of all  Seyfert galaxies observed with  BeppoSAX included 68 objects among 18
Sy 1, 10 Sy 1.5 and 40 Sy 2 galaxies.\\
Our sample is finally composed of 67 observations for a total of 45 Seyfert galaxies of which  14 are Seyfert 1 galaxies, 9 are Seyfert 1.5
and 22 are  Seyfert 2 galaxies. \\
The list of objects composing our subsamples, their characteristics and their spectral properties
is given in  Table A.1. for Sy 1 and Sy 1.5 galaxies samples and in  Table A.2. for the Sy 2 galaxies
sample. 
\section{Data Analysis}
The BeppoSAX instruments cover a wide spectral band and consist of four co-aligned
Narrow Field Instruments (NFI) plus two Wide Field Cameras perpendicular to the axes of the NFI and
looking in opposite directions.\\
The NFI is composed by a Low Energy Concentrator Spectrometer (LECS: \citet{Parmar}), three
Medium Energy Concentrator Spectrometer (MECS: \citet{Boella}) and a   Phoswich Detector
Counter (PDS: \citet{Frontera}). LECS and MECS have imaging capabilities and operate
respectively in the 0.1-5 keV and 2.0-10 keV. The instrument used in this study, the PDS, is an instrument using rocking
collimators, the energy band is between 15-200 keV. The PDS consists in four phoswich units 
operating in collimator mode with two of them pointing towards the source,  the other two 
away.  The two pairs switch on and off source and the net source spectra are 
obtained by subtracting the 'off' (which represent the background) from the "on" counts.
The PDS data available on the web are already background subtracted. 
A more detailed description of the background and of the subtraction procedure is available
from \textit{http://www.sdc.asi.it/software}.  \\
The data analysis has been performed with the XSPEC \citep{Arnaud} version 11 package and using the
last PDS response matrices
released by the SAX Data Center.\\
We kept in the input and output files the same binning for the average spectrum as was
available for the
original data of each source. 
For a source observed several times, we first compute its 
average spectrum. 
We then compute the average subclass spectrum  
taking into account of the exposure time of each data set.
The energy band is from 15  to 136 keV. 
\section{Spectral properties of the different classes of Seyfert galaxies}  
We present in Figure 7  the average counts spectrum of the three classes of Seyfert galaxies.

\begin{figure}[!hbt]
\centering
\includegraphics[width=6.5cm, angle=-90]{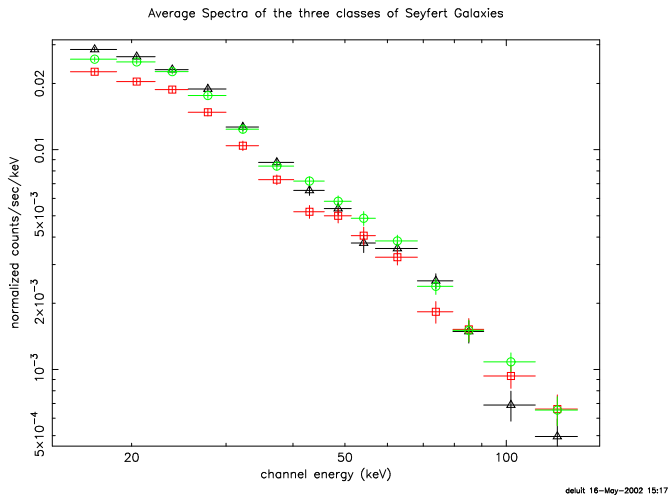}
\caption{Average spectra of the three classes of Seyfert galaxies: Sy 1 (triangle), Sy 1.5 (square) and Sy 2 galaxies
(circle)}
\end{figure}

The spectrum of each class  has an \textit{integrated} signal to noise ratio
 of  96$\sigma$
 for Seyfert 1 galaxies, 74.4$\sigma$  for Seyfert
1.5 galaxies and 94$\sigma$ for Seyfert 2 galaxies. 
No normalization has been applied to obtain the average spectra shown in this figure.\\
We compare the Seyfert classes following two ways. First through a model  independent  method comparing
 directly the counts spectra. Secondly  
 through a model dependent method using a fitting procedure to investigate the physical processes at the origin
  of the emission.
\begin{figure*}[!hbt]
\centering
\includegraphics[width=12cm]{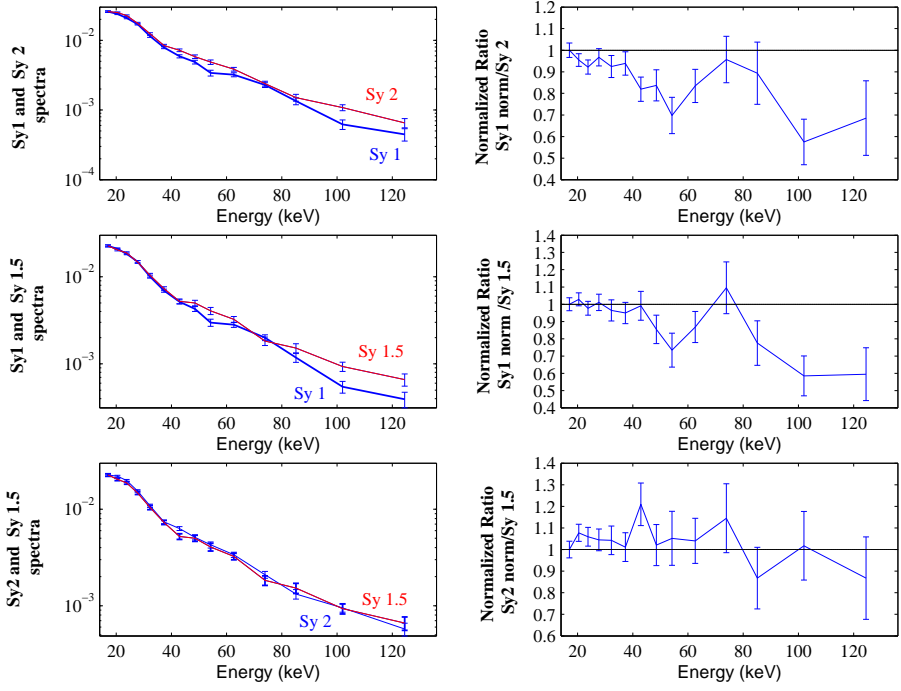}
\caption{Left: three pairs of Spectral Energy Distributions normalized at 15 keV. Right: ratios of these SEDs.}
\end{figure*}
\subsection{Model Independent Method: comparison of the normalized counts spectra}
We compare the different average spectra  normalizing them  at 15 keV. 
This allows us to compare their emission independently of the average flux. The result is shown in Figure 8. 
 On the left column,  we present the normalized counts spectra of the three pairs of  classes, on the right side we 
 give the ratio of each pair. 
\subsubsection{Comparison between Seyfert 1 and Seyfert 2  emission}
The two spectra are similar
 below 40 keV confirming that the Sy 2 galaxies emission  is not significantly modified by absorption effects.  
Above 40 keV, the Sy 2 emission dominates that of Sy 1. The difference between the
 two classes increases above 90 keV. Indeed, the hard photons population is more important in Sy 2 than in
  Sy 1 galaxies showing that the \textit{observed} Sy 2 spectrum is harder than Sy 1. The deficit of hard photons
   in the Sy 1 class can be explained either by the presence of a cutoff in its emission and/or by a reflection process
    occurring in the Sy 2 emission. The possibility that the Sy 2 \textit{intrinsic} power law  is harder than
     Sy 1 cannot be excluded. The Sy 1 average emission  is 42\%  weaker than that of Sy 2 around 100 keV where they
      differ most in the normalized ratio.
\subsubsection{Comparison between Seyfert 1 and Seyfert 1.5 emission}
The spectra are identical below 40 keV. The Sy 1.5 emission slightly dominates between 40 and 65 keV. At higher 
energies, the Sy 1.5 spectrum contains a more important hard photons 
 population. As in Sy 2
  galaxies, the  Sy 1.5 emission domination above 80 keV  can be produced by a cutoff in Sy 1 or by a harder Sy 1.5
   primary 
     power law.  
The Sy 1 average emission  is 41\%  weaker than that of Sy 1.5 at $\sim$100 keV where they
      differ most in the normalized ratio.
\subsubsection{Comparison between Seyfert 1.5 and Seyfert 2 emission}
The two  spectra remain very close  along the PDS energy band.  Indeed, the maximal deviation
 between the two spectra in the normalized ratio does not exceed 20\% at $\sim$40 keV  where the Sy 2 average
emission  is marginally higher than  that 
 of Sy 1.5.   
  Thus, we consider 
in a first approach that the spectrum of Sy 1.5 and Sy 2 present notable similarities. 
\subsubsection{Summary}
The main result of this model independent study  is that Sy 2 differ from Sy 1 galaxies and that Sy 1.5 and Sy 2 are 
 similar.  
 In particular in the high
 energy domain where the Sy 1.5 and Sy 2 classes present a more important hard photons population compared with Sy 1
  galaxies. The processes that can be responsible for the
    observed differences between the three classes is investigated in the next subsection.
\subsection{Model Dependent Method: investigating the physical processes into Seyfert
galaxies engine}
To interpret our data in terms of models taking the PDS spectral response matrices into account, we
  fit
  the spectra with different theoretical models. 
  We first fit the spectra with simple models (power law, broken power law and cutoff
  power law). We then  consider  reflection  with
   the PEXRAV model described below.\\
  All the results are summarized in  Tables 1 and 2.
 \begin{table*}[!hbt]
 \begin{center}
   \caption{Best fit parameters for  simple models}
\begin{tabular}{c c c c c c c c}
\hline
\hline
 Model & Seyfert & $\Gamma$ & E$_{break/cutoff}$ & $\Gamma_{2} $ & $\chi^{2}$/d.o.f. &   $\chi^{2}_{r}$ & FTEST \\
 & Class  & Index Slope& (keV)&Index Slope &  & reduced& Probability value \\
\hline
Power Law & 1   & 2.03 & - & - & 50.0/12 & 4.17 & - \\
  & 1.5  & 1.90$_{-0.05}^{+0.05}$   & - & - & 16.7/12 & 1.39&  - \\
 & 2  & 1.90  & - & - & 44.2/12 & 3.68 &  -\\
\hline
Broken Power Law & 1 & 1.65$_{-0.21}^{+0.16}$ & 28$_{-3.8}^{+7.0}$ & 2.25$_{-0.09}^{+0.12}$ &
14.7/10 & 1.47 & 2.2$\cdot 10^{-3}$ \\
 & 1.5 & 1.62$_{-0.38}^{+0.21}$ & 26$_{-5.4}^{+16}$ & 2.01$_{-0.09}^{+0.13}$ & 10.3/10 & 1.03 & 5.0$\cdot$10$^{-2}$ \\
 & 2 & 1.36$_{-0.56}^{+0.24}$ & 24$_{-2.1}^{+4.3}$ & 2.04$_{-0.06}^{+0.07}$ &
13.9/10 &1.39 & 3.1$\cdot$10$^{-3}$ \\
\hline
Cutoff Power Law & 1 & 1.49$_{-0.17}^{+0.17}$ &
70$_{-17}^{+30}$ & - &
15.2/11 & 1.38 & 4.5$\cdot$10$^{-4}$ \\
   & 1.5 & 1.64$_{-0.19}^{+0.11}$ & 152$_{-66}^{+48}$ & - &
13.3/11 & 1.21 & 5.8$\cdot10^{-2}$ \\
 & 2 & 1.45$_{-0.17}^{+0.15}$ & 87$_{-24}^{+42}$ & - &
17.6/11 & 1.60 & 1.8$\cdot10^{-3}$ \\
\hline
\end{tabular}
\end{center}
Note:\\
 The uncertainties correspond to 90$\%$
confidence level based on a $\Delta$$\chi^{2}$=2.7 criterion \citep{Lampton}\\
\end{table*}
\begin{table*}[!hbt]
\begin{center}
 \caption{Best fit parameters for the reflection model PEXRAV}
\begin{tabular}[!hbt]{c c c c c c c}
\hline
\hline
\multicolumn{7}{c}{PEXRAV PARAMETERS}\\
\hline
Seyfert & $\Gamma$ & Efolded & R  & $\chi^{2}$/d.o.f. &  $\chi^{2}_{r}$& cos $\theta$  \\
Class & Index Slope & (keV) & Reflection & & reduced& fixed   \\
\hline
 1 & 1.92$_{-0.49}^{+0.26}$ & 238$_{-176}^{+\textrm{\tiny{NC}}}$ &
 0.93$_{-\textrm{\tiny{NC}}}^{+2.00}$ &13.2/10 & 1.32 & 0.45  \\
 1.5 & 1.96$_{-0.06}^{+0.06}$ &  9.8$\cdot$10$^{5}  _ {\tiny{[a]}}$ & 0.73$_{-0.71}^{+0.81}$  & 10.8/10 & 1.08 &  0.45  \\
  2 & 1.99$_{-0.45}^{+0.05}$ & 2.4$\cdot$10$^{4} _ {\tiny{[a]}}$ & 1.47$_{-0.77}^{+1.07}$ & 13.1/10 & 1.31 & 0.45  \\
\hline
 1 & 1.89$_{-0.46}^{+0.28}$ & 221$_{-158}^{+\textrm{\tiny{NC}}}$ &
0.48$_{-\textrm{\tiny{NC}}}^{+1.54}$ & 13.1/10&
1.31
&  0.87  \\
   1.5 & 1.95$_{-0.05}^{+0.06}$ & 8.2$\cdot$10$^{5} _ {\tiny{[a]}}$ & 0.39$_{-0.26}^{+0.38}$ & 10.7/10 & 1.07 & 0.87  \\
   2& 2.00$_{-0.45}^{+0.05}$ & 3.6$\cdot$10$^{4} _ {\tiny{[a]}}$ &
   2.33$_{-0.49}^{+1.96}$ & 13.0/10 &
 1.30 & 0.30 \\
\hline
Seyfert & $\Gamma$ & Efolded & R  & $\chi^{2}$/d.o.f. &  $\chi^{2}_{r}$&  cos $\theta$ \\
Class  & Index Slope & (keV) & Reflection & & reduced & free   \\
 \hline
 1 & 1.91$_{-0.47}^{+0.29}$ & 231$_{-168}^{+\textrm{\tiny{NC}}}$ &
 0.53$_{-0.27}^{+\textrm{\tiny{NC}}}$ & 13.2/9 & 1.47 & 0.81  \\
   1.5 & 1.95$_{-0.06}^{+0.09}$ & 1.0$\cdot$10$^{6} _ {\tiny{[a]}}$ &
   0.36$_{-0.23}^{+\textrm{\tiny{NC}}}$ & 10.7/9  & 1.19 & 0.95 \\
   2& 2.00$_{-0.44}^{+0.05}$ & 2.8$\cdot$10$^{3} _ {\tiny{[a]}}$ &
   2.55$_{-2.13}^{+\textrm{\tiny{NC}}}$ & 13.0/9 & 1.44 & 0.28  \\
\hline
    \end{tabular}
\end{center}

   Notes:\\
The uncertainties correspond to 90$\%$
confidence level based on a $\Delta$$\chi^{2}$=2.7 criterion \citep{Lampton}.\\
\textbf{NC} indicates that no constraint has been found on this limit.\\
\textbf{[a]} the value indicates that the high energy cutoff is far above values that can be measured
by the 
PDS showing the absence of a measured cutoff.
\end{table*}
 \subsubsection{Fitting of the spectra with simple theoretical models}
 The power law model gives
 the worst statistical results for all the classes. For Sy 1 and 2 galaxies, this model is
 statistically unacceptable with a reduced $\chi$$^{2}$ greater than 2 indicating that their spectra require
  an additional component.  For the Sy 1.5 galaxies, this model
 is  more 
 adapted.\\ 
 Sy 1 emission  requires a
 cutoff considering
 both the statistical result of the $\chi^{2}$ test and the FTEST probability,   
confirming our model independent analysis. The presence of a cutoff
 is  unprobable in Sy 1.5 and Sy 2 galaxies. Indeed, the $\chi$$^{2}$ method gives the best result for
  a broken power law ($\chi^{2}_{r}$=1.03 for Sy 1.5 and $\chi^{2}_{r}$=1.30 for Sy 2) compared with the cutoff
   model ($\chi^{2}_{r}$=1.21 and $\chi^{2}_{r}$= 1.60 respectively). \\ 
In subsection 5.1,  we noted two maxima in the ratio between Sy 1 and Sy 2 spectra around 40 keV and 100 keV. These
features may be due to a combination of a cutoff in Sy 1 (here found at 70$_{-17}^{+30}$ keV) and/or a reflection hump in  Sy 2 class that we
investigate in the next section.
\subsubsection{Fitting of the spectra with the reflection model PEXRAV}
Previously, we suspected the presence of a reflection component in the Sy 1.5 and Sy 2 spectrum. We investigate this hypothesis using the PEXRAV model \citep{Magdziarz} in XSPEC. This model
calculates the expected X-ray spectrum when a point source of X-rays is incident on optically thick,
mainly neutral material. 
The parameter R  measures the reflection
component. If the primary
source is isotropic and neither the source nor the reflector are obscured, then R can be linked to the
solid angle seen through the reflector as R$\simeq$$\frac{\Omega}{2\pi}$. 
This model and in particular the reflection component are  inclination dependent. 
We  apply this model in two ways: first, we fix the inclination from the normal of the disk at a medium
 angle  to test
 the occurrence of the  reflection as well as the presence of other components independently to any 
 strong inclination influence. We then  change the angle value fixing or letting it free in order 
 to understand its role  and to obtain the inclination value for which we have 
the best fit. The results are shown in  Table 2.
\begin{enumerate}
\item{Intermediate inclination analysis} \\ 
The inclination is  initially fixed at an intermediate value of cos $\theta$=0.45.\\
The $\chi^{2}$ test indicates that PEXRAV significantly improves the data fitting of Sy 1 and Sy 2 galaxies compared
 with simple models, not however for the Sy 1.5 spectrum that remains best represented by a broken power law. 
  Therefore,  reflection may play a role for the two main Seyfert classes and be absent in  
 Sy 1.5 emission. This is confirmed by the reflection amount which can be non existent for
  Sy 1.5 (R=0.73$_{-0.71}$) whereas its presence is strongly suggested  in Sy 2 emission (R=1.47$_{-0.77}$)  at the 2$\sigma$ level.
The primary power law indices  of the three classes are within 1$\sigma$, however Sy 2 emission could be softer
 compared with the Sy 1
 index slope ($\Gamma$=1.99$_{-0.45}^{+0.05}$  for Sy 2 against $\Gamma$=1.92$_{-0.49}^{+0.26}$  for Sy 1). \\
A  strong difference appears in the high energy cutoff which is relatively well identified
 for Sy 1 (E$_{c}$=238$_{-176}$ keV) and totally absent for Sy 1.5 and Sy 2 classes. 
 The main result is that the spectrum of Sy 2 galaxies differs from that of Sy 1 
 with  the absence
of a cutoff and a possible presence of an important reflection component in the Sy 2 emission.  
\item{Inclination influence}\\
We change the inclination value  to test  unified models which suggest that Sy 2 galaxies are seen
edge-on.
\begin{enumerate}
\item Fixed inclination at different angles\\
We fix the inclination at cos $\theta$=0.30 and cos $\theta$=0.87. The influence of the
 inclination parameter is not crucial on the final result of the $\chi^{2}$ test.
The best fit for Sy 1 and Sy 1.5 is obtained for  
cos $\theta$=0.87, an angle of 30$^{\circ}$ from the normal of the disk. The reflection component
 value is lower for cos $\theta$=0.87 than for an inclination of cos $\theta$=0.45 as predicted by the fact
  that the reflection is inclination dependent in the PEXRAV model.
We also confirm that we obtain for the Sy 2 galaxies the best $\chi^{2}$ value for   cos $\theta$=0.30
namely  an angle of 73$^{\circ}$. The increase of the reflection component
is numerically important (2.33 against 1.47 previously) but statistically  unconvincing. The reflection  is less
predominant in the two other Seyfert classes.
\item Free inclination \\
We then let the inclination free to avoid as much as possible the dependence between the different parameters.     
We find that Sy 1.5 galaxies would be seen closer  to the normal to the disk with an inclination of 
18$^{\circ}$ whereas the Sy 1 class
 would be viewed from an angle of 36$^{\circ}$ and the Sy
2  from 74$^{\circ}$.\\
It should be noted that in the PEXRAV implementation of the reflection model, a stronger reflection
hump systematically induces a large inclination.
\end{enumerate}
\end{enumerate}

\subsection{Investigating a link between the index slope $\Gamma$ and the reflection parameter R}
We investigate  whether the index slope $\Gamma$ and the reflection R are independently measured  through the "draw" of 
this possible correlation using XSPEC (Figures 9 and 10).

\begin{figure}[!h]
\centering
\includegraphics[width=6cm, angle=-90]{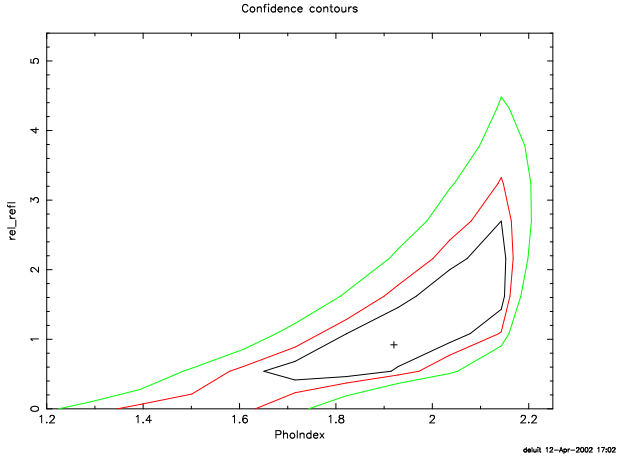}
\caption{The 68\%, 90\% and 99\% confidence contours of the index slope $\Gamma$ and the Reflection
parameter R for the Sy 1 average spectrum.}
\end{figure}
\begin{figure}[!h]
\centering
\includegraphics[width=6cm, angle=-90]{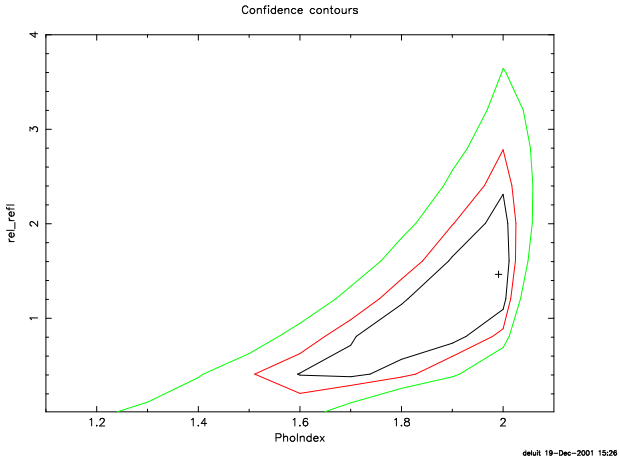}
\caption{The 68\%, 90\% and 99\% confidence contours of the index slope $\Gamma$ and the Reflection
parameter R for the Sy 2 average spectrum.}
\end{figure}

This study is performed using an intermediate angle of cos  $\theta$=0.45
 for all classes. 
For both Sy 1 and Sy 2 classes, if the average spectrum has a high index slope (e.g. $\Gamma$= 2) then it 
 has  a larger  amount of reflection than if it has a hard spectrum (e.g. $\Gamma$=1.6).  
We also note that if any class would have an index slope of less than 1.2, then reflection is  not  required   
to account for the data. \\
  The index slope is the signature of the
 primary X-ray source. Therefore,  a link  between the index slope and the reflection amount  
  would indicate 
  that the source of the cooling photons is related to the emission from the  material at the origin of the 
  Compton reflection.
   We will study using individual object spectra whether there is a correlation between $\Gamma$ and R in the Paper
    II \citep{Deluit2}.
  
\subsection{Results}
With a model independent method we find that Sy 2 galaxies are  similar to Sy 1.5 and differ
 from Sy 1 galaxies. \\
 With a model dependent method, the best of all theoretical models adopted for the
two main  classes is the PEXRAV model whereas the Sy 1.5 spectrum is well represented   by a broken power
 law. \\
  Considering the cutoff power law model, E$^{-\Gamma}$exp(-E/E$_{c}$), we find that Sy 1.5 and
 Sy 2 have similar behavior in the sense that their spectra do not require any cutoff 
 whereas the Sy 1 spectrum presents a well constrained cutoff at 70$_{-17}^{+30}$ 
 keV. In model including reflection, the cutoff energy in Sy 1 emission is found at 221$_{-158}$ keV and still not
 required in Sy 1.5 and Sy 2 emission.\\
The main result is thus the presence of a high energy cutoff in Sy 1 that it is absent in
 other categories. The role and place of Sy 1.5 are not clearly identified. Indeed, Sy 1.5 are close to Sy 1
  concerning the relatively low reflection 
influence,  and similar  to Sy 2
regarding the  absence of a cutoff. 
\subsection{Discussion and Perspectives}
Our results concerning the index slope of Sy 1 emission confirm the results obtained by \citet{Z95} and \citet{Gondek}
 who 
found
$\alpha$$\sim$0.9. Our results differ, however, in what regards Sy 2 galaxies. \citet{Z95} and \citet{Z2000} found
 that Sy 2 are harder than Sy 1 which we do not find. Secondly, \citet{Z95} found a negligible
  reflection in Sy 2 emission whereas in our study, we emphasize the possible importance of this
  component in
  the Sy 2
average spectrum. 
\subsubsection{High energy cutoff}
The presence or absence of a high energy cutoff is not influenced by the inclination of the system with
respect to the line of sight. The measured cutoff in the Sy 1 galaxies emission while none is found in Sy 2 galaxies
  clearly indicates intrinsic differences in their primary emission. \\
However this lack could also be explained by a wide  
 distribution of  cutoff energies whereas the presence of a cutoff in the Sy 1 average spectrum indicates a narrow
 range of comptonizing temperatures in the class. \\
A cutoff in the Sy 1 emission provides, according to thermal Compton models, a reliable
estimate of the electron temperature in the Comptonizing medium. The Sy 1 cutoff found at 
 k$_{B}$T=220 keV corresponds
 to a temperature of 
T$\sim$2.6$\cdot$10$^{9}$
 K for  the electron gas in the case of unsaturated comptonization. \\
If the absence of a cutoff in Sy 1.5 and Sy 2 galaxies is confirmed, it  indicates either the 
importance of non thermal Compton
scattering or a higher  temperature of the Comptonizing medium.	   
\subsubsection{The role of reflection}
We show that Sy 2 galaxies may have a stronger reflection component than the other classes. Our
fits also indicate that the inclination of Sy 2 could be larger than that of the other galaxies. Both
results are not independent, only models free of this degeneracy will allow us to deconvolve
the geometry of the reflection medium from inclination effects. \\
We  also find that the average spectrum of Sy 2 galaxies, more affected by absorption than Sy 1, shows more reflection
than that of Seyfert 1 galaxies suggesting a link between the two processes.
\subsubsection{Implications of our results}
To investigate further these assertions, it is necessary to test our results
  in a different way, through the study of the distribution of the individual spectral
   parameters on a large sample of Seyfert galaxies.  Therefore, we will study the distribution  of the
 parameters $\Gamma$, E$_{c}$ and R for each object of a given class as far as can be measured to date and to compare 
them in order to test the results found in this paper. This study, presented in  Paper II, will
 allow us  to know
 whether the absence of a high energy cutoff
  in Sy 1.5 and Sy 2 is due to a wide distribution of values within the class and to study the correlation between
  the reflection
  and the  power law slope.\\ 
Seyfert 2 galaxies have different intrinsic emission and  therefore  different physical conditions in the inner regions. 
Sy 2 galaxies in which no cutoff has been found have probably a higher temperature
       comptonizing medium
      than Sy 1 galaxies. This reveals that the ratio heating/cooling of this medium is different in both classes.\\
      Our study has to be pursued within the Sy 2 galaxies where two subclasses known as Sy 2 with Polarized Broad Lines 
      (PBL) 
and without PBL are found. Unified schemes predict that both types have a hidden  Sy 1 nucleus. We have to search for common
 behavior in the intrinsic emission of both Sy 2 subclasses and to compare them with Sy 1 emission features
  (see \citet{DeluitPBL}).  \\
The nature of
  the fueling mechanism occurring in the nucleus is unsolved and could also contribute to the  
   Seyfert 2  particularity. Indeed,  Sy 2 galaxies have been found to host starbursts more often than type 1 AGNs
   \citep{Levenson2001}. We could expect that in Sy 2 objects the accreted  matter 
   originates predominantly in starbursts.        
 \section{Conclusion} 
We obtained and studied the average hard X-ray spectra of Seyfert 1, 1.5  and 2 galaxies  
observed by the PDS on board BeppoSAX. 
 We emphasized several differences in the central engine behavior of the two main Seyfert classes 
indicating  that the Seyfert 1 and 2 galaxies have at least various physical conditions in their inner regions or are 
  different objects. The differences cannot be due only to inclination/absorption effects as claimed by unified models. \\
The next  steps in the Seyfert studies will be the search for the presence of a high energy cutoff
 in individual Seyfert galaxies to investigate the importance of the different physical processes in their emission
  (thermal and/or non thermal) and the investigation of the nature of the reflection medium
   (disk or material surrounding the nucleus), of its composition (density, homogeneity)  and its localization. 
 It is also important in a different set of considerations to investigate whether Seyfert 1 and Seyfert 2 have intrinsic differences in their
star formation rate, morphology and/or environment.
\bibliographystyle{apj}
\bibliography{biblio}

\appendix
\onecolumn
\newpage
\section{The final sample of the objects studied}
We present below the characteristics of the objects composing the final sample of  our study. The Table A.1. includes the Seyfert 1 and Seyfert 1.5 galaxies subsamples and the Table A.2. represents the Seyfert 2 subsample. 
\begin{table*}[!hbt]
\begin{center}
\caption{General Characteristics and Spectral Properties of the objects composing the Sy 1 and Sy 1.5 samples}
\begin{tabular}[!hbt]{c c c c c c c}
\hline
\hline
\large{Source Name} &
\large{RA}   &
\large{DEC}  &
\large{Redshift} &
\large{S/N}$^{1}$ &
\large{F$_{\tiny{15-136 ~\textrm{keV}}}$}$^{6}$ &
\large{L$_{\tiny{15-136 ~\textrm{keV}}}$}$^{7}$ \\
  & (\it{h  m  s})& ($^{\circ}$  $^{\prime}$  $^{\prime\prime}$)  & z & & &  \\
\hline
\multicolumn{7}{c}{\large{Seyfert 1 galaxies sample}}\\
\hline
Mrk 335 & 00 06 19.3 & +20 12 10.0 & 0.025 & 3.01$\sigma$ & 1.38 & 1.67 \\   
Fairall 9 & 01 23 45.6 & $-$58 48 14.0 & 0.047 & 11.8$\sigma$ & 5.42 & 23.5  \\
NGC 985 & 02 34 37.6 & $-$08 47 08.0 & 0.043 & 9.69$\sigma$  & 3.63 & 13.1 \\
1H0419-577 & 04 26 01.0 & $-$57 12 00.0 & 0.104 &  2.00$\sigma$ & 1.18 & 25.7 \\
Mrk 110$^{2}$ & 09 25 12.9 & +52 17 11.0  &0.035 & 12.8$\sigma$ & 4.35 & 10.4  \\
REJ1034+393 & 10 34 38.8 & +39 38 34.0 & 0.042 & 2.14$\sigma$ & 0.71 & 2.44 \\
NGC 3783 & 11 39 01.7 & $-$37 44 20.0 & 0.011 & 48.5$\sigma$ & 14.2 & 3.31 \\
NGC 4051$^{2}$  & 12 03 09.5 & +44 31 50.9 & 0.002 & 7.25$\sigma$ & 2.06 & 0.02\\
NGC 4593 & 12 39 39.4 & $-$05 20 38.0 & 0.008 & 23.0$\sigma$ & 9.63 & 1.19 \\
MCG 6-30-15 & 13 35 53.2 & $-$34 17 48.0 & 0.008 & 33.0$\sigma$ & 7.19 & 0.84 \\
IC 4329A$^{5}$ & 13 49 19.2 & $-$30 18 35.9 & 0.016 & 99.0$\sigma$ & 22.6 & 11.2 \\
ESO 141-G55 & 19 21 14.3 & $-$58 40 13.0 & 0.036 & 15.3$\sigma$  & 5.37 & 13.6 \\
Mrk 509$^{2}$ & 20 44 09.7 & $-$10 43 23.9 & 0.034 & 27.4$\sigma$ & 9.73 & 21.9 \\
NGC 7469 & 23 03 15.5 & +08 52 26.0 & 0.016 & 28.0$\sigma$  & 5.11 & 2.52\\
\hline
\multicolumn{7}{c}{\large{Seyfert 1.5 galaxies sample}}   \\
\hline
Mrk 1152 & 01 13 50.2 & $-$14 50 59.9 &0.053 & 3.32$\sigma$ & 0.59 & 3.22 \\
NGC 526A  & 01 23 54.1 & $-$35 03 56.0 &0.019 & 13.7$\sigma$  & 5.89 & 4.15 \\
MCG 8-11-11 & 05 54 53.5  & +46 26 21.0 &0.021 & 29.3$\sigma$ & 11.4 & 9.25 \\
Mrk 6 & 06 52 12.1 & +74 25 36.9  &0.019 & 21.0$\sigma$  & 7.67 & 5.24 \\
NGC 3516$^{2}$ & 11 06 47.2 & +72 34 08.0   & 0.009 & 29.8$\sigma$ & 11.5 & 1.71 \\
Mrk 766/NGC4253$^{2}$ & 12 18 26.2  &  +29 48 46.0  &0.013 & 14.0$\sigma$ & 3.63 & 1.16 \\
NGC 5548$^{5}$ & 14 17 595 & +25 08 12.0 & 0.017 & 52.6$\sigma$ & 10.6 & 6.06 \\
Mrk 841 & 15 04 01.2& +10 26 16.0  & 0.036& 7.71$\sigma$ & 2.71 & 6.92\\
NGC 7213$^{3}$& 22 09 16.2 & $-$47 10 00.0    &0.006 & 31.4$\sigma$ & 7.03 & 0.49 \\
\hline 
\end{tabular}
\end{center}
 Note:\\
$^{1}$ Integrated Signal to Noise ratio in the PDS energy range\\
$^{2}$ 2 available observations \\
$^{3}$ 3 available observations\\
$^{4}$ 4 available observations\\
$^{5}$ 5 available observations\\
$^{6}$ Integrated Flux in 10$^{-11}$erg\,s$^{-1}$\,cm$^{-2}$\\
$^{7}$ Integrated Luminosity in 10$^{43}$erg\,s$^{-1}$\\
\end{table*}

\begin{table*}[!hbt]
\begin{center}
\caption{General Characteristics and Spectral Properties of the objects composing the Sy 2 sample}
\begin{tabular}{c c c c l c c c }
\hline
\hline
\large{Source name} &
\large{RA}   &
\large{DEC}  &
\large{Redshift}	&
\large{~~~N$_{\footnotesize{\textrm{H}}}$}$^{1}$ &
\large{S/N}$^{4}$ &
\large{F$_{\tiny{15-136 ~\textrm{keV}}}$}$^{5}$ &
\large{L$_{\tiny{15-136 ~\textrm{keV}}}$}$^{6}$  \\
  &(\it{h  m  s})& ($^{\circ}$  $^{\prime}$  $^{\prime\prime}$)   & z &  & & & \\
  \hline
IRAS 00198-7926 & 00 21 53.8 & $-$79 10 08.0 & 0.073 &~~~~~~-  & 2.10$\sigma$ & 2.18 &
22.9 \\
NGC 1358 & 03 33 39.5 & $-$05 05 20.0  & 0.013 &~~~~~~- &2.28$\sigma$ & 1.38 & 0.48 \\
IRAS 05189-2524 & 05 21 01.3 & $-$25 21 42.9  & 0.043& ~~4.90\tiny{$^{[B]}$}  &  2.66$\sigma$ & 1.14 & 4.04\\
NGC 2110 & 05 52 11.2 & $-$07 27 20.8  & 0.008 & ~~2.89\tiny{$^{[H]}$}  & 20.5$\sigma$ & 8.62 & 1.01 \\
NGC 2992$^{2}$ & 09 45 42.0 & $-$14 19 36.9 & 0.008& ~~0.69\tiny{$^{[W96]}$}  & 25.8$\sigma$ & 7.98 & 0.91 \\
MCG 5-23-16 & 09 47 40.1 & $-$30 56 53.9 & 0.008& ~~1.62\tiny{$^{[W97]}$} &45.8$\sigma$ & 19.8 & 2.62 \\
IRAS F12072-0444 & 12 09 45.1 & $-$05 01 14.9 & 0.128&~ 0.17\tiny{$^{[D]}$} & 
4.88$\sigma$ & 1.84 & 61.5  \\
NGC 4388$^{2}$& 12 25 46.7 & +12 39 44.0 & 0.008&~~42.0\tiny{$^{[B]}$} &  63.8$\sigma$ & 22.1 &
3.00\\
NGC 4507$^{3}$ & 12 35 37.0 & $-$39 54 32.0 & 0.012 & ~~29.2\tiny{$^{[C]}$} & 58.1$\sigma$ &
19.1 & 5.12 \\
NGC 4939 & 13 04 14.5 & $-$10 20 26.9  & 0.010 &  ~~30.0\tiny{$^{[M]}$} &  3.56$\sigma$ & 2.26 & 0.47  \\
NGC 5252 & 13 38 16.3 & +04 32 20.0  & 0.023 &~~4.33\tiny{$^{[Ca]}$}    & 2.86$\sigma$ & 2.26 & 2.34\\
NGC 5506$^{3}$ & 14 13 14.8 & $-$03 12 28.0 & 0.006  &~~3.40\tiny{$^{[S]}$}   & 69.0$\sigma$ & 17.0 & 1.25 \\
NGC 5674 & 14 33 52.0 & +05 27 28.0 & 0.025 &~~7.00\tiny{$^{[B]}$} & 5.05$\sigma$ & 1.69 & 2.07  \\
NGC 6300 & 17 17 00.3 & $-$62 49 15.0 & 0.004 & ~~60.0\tiny{$^{[Le]}$}     & 23.1$\sigma$ & 8.35 & 0.22  \\
IRAS 18325-5926/F 49 & 18 36 58.3 & $-$59 24 09.0 & 0.020 &~~1.26\tiny{$^{[B]}$} & 6.06$\sigma$ & 1.25 & 0.99 \\
ESO 103-G35 & 18 38 20.3 & $-$65 25 40.0 & 0.013 &~~15.9\tiny{$^{[T]}$}  & 7.84$\sigma$ & 7.96 &
2.71 \\
IRAS 20210+1121 & 20 23 25.6 & +11 31 32.9 & 0.056 &~~$\leq$6.0\tiny{$^{[U]}$} & 2.24$\sigma$ & 0.87 & 5.45 \\
NGC 7172$^{2}$& 22 02 02.2 & $-$31 52 12.0 & 0.009 &~~8.61\tiny{$^{[G]}$} &
9.73$\sigma$ & 4.43
& 0.65 \\
NGC 7314 & 22 35 46.0 & $-$26 03 01.0 & 0.005 &~~1.16\tiny{$^{[T]}$}   & 18.5$\sigma$ & 7.07 & 0.30 \\
IRAS 23060+0505 & 23 08 33.8 & +05 21 29.0 & 0.173 & ~~8.40\tiny{$^{[Br]}$} & 4.20$\sigma$ & 1.79 &
112 \\
NGC 7582 & 23 18 23.5 & $-$42 22 11.9  & 0.005 & ~~12.4\tiny{$^{[X]}$}& 24.2$\sigma$ & 10.5 & 0.57\\
NGC 7679 & 23 28 46.8 & +03 30 38.9  & 0.017   &~~0.04\tiny{$^{[DC]}$} & 4.98$\sigma$ & 1.36 & 0.76 \\
\hline
\end{tabular}
\end{center}

Note:\\
\textit{References concerning the Hydrogen column density:} \\
	B: \citet{Bassani}, 
	Br: \citet{Brandt},  
	Ca: \citet{Cappi}, 
	Co: \citet{Comastri} 
	D: Deluit, this work, see Paper II of this study, 
	DC: \citet{DC}, 
	G: \citet{Guainazzi98},  
	H: \citet{Hayashi}, 
	Le: \citet{Leighly}, 
	M: \citet{Maiolino}, 
	S: \citet{SmithDone} , 
	T: \citet{Turner}, 
	U: \citet{Ueno}, 
	W96: \citet{Weaver96},  
	W97: \citet{Weaver97}, 
	X: \citet{Xue} \\

$^{1}$ Hydrogen column density in 10$^{22}$ cm$^{-2}$ units\\
$^{2}$ 2 available observations \\
$^{3}$ 3 available observations\\
$^{4}$ Integrated Signal to Noise Ratio\\
$^{5}$ Integrated Flux in 10$^{-11}$erg\,s$^{-1}$\,cm$^{-2}$\\
$^{6}$ Integrated Luminosity in 10$^{43}$erg\,s$^{-1}$\\
 
\end{table*}
\newpage 
\section{The role of absorption in Seyfert 2 class studies } 
According to actual models, the X-ray emission of many Seyfert 2 galaxies is characterized by a power law
similar to that observed in Seyfert 1, with a cutoff at low energies ($<$ 10 keV) due to absorption
effects produced by gas column densities higher than 10$^{22}$cm$^{-2}$. Seyfert 2 galaxies show a broad
 range of absorption, with column densities varying from $\leq$ 10$^{22}$
cm$^{-2}$ to
 $\geq$ 10$^{25}$ cm$^{-2}$ \citep{Risaliti}. These results
  yield to a further
  classification
  among Seyfert 2 galaxies: Compton thin and Compton thick sources depending on whether their column 
  density is respectively lower 
  or larger
   than the inverse Thomson cross-section 
   N$_{\footnotesize{\textrm{H}}}$= $\sigma_{\footnotesize{\textrm{T}}}^{-1}$ =1.5$\cdot$10$^{24}$
   cm$^{-2}$. \\
Therefore, for objects presenting a column density 
less than  1.5$\cdot$10$^{24}$cm$^{-2}$
(\textit{Compton thin}), high energy X-rays  penetrate the absorbing material, making the
nuclear source visible to the observer and allowing him to measure the Hydrogen column density. For objects with a column density higher than
1.5$\cdot$10$^{24}$cm$^{-2}$ (\textit{Compton thick}), the matter is optically thick to the Compton scattering
and the nucleus is hardly visible. 
In the last few years, BeppoSAX data revealed a new highly absorbed class with only a lower limit on the column
density  N$_{\footnotesize{\textrm{H}}}$$>$ 10$^{25}$cm$^{-2}$ \citep{Risaliti}, no detection of the intrinsic emission can
be  expected at energies between 15 and 150 keV.\\
The photoelectric absorption cross section as a function
 of
 energy is  well described by
\citet{MMC} up to 10 keV.
The component due to Compton scattering  is described by the Thomson cross
section equal to 6.65$\cdot$10$^{-25}$cm$^{-2}$.
Compton scattering plays an important role with the photoelectric cross section above 4 keV, and
dominates above 10 keV. Starting with the analytic formulae \citet{MMC}
\begin{equation}
\sigma_{photoelectric}= (C_{0} + C_{1}E + C_{2} E^{2})E^{-3}\times10^{-24} \textrm{cm}^{-2},
\end{equation}
where C$_{0}$=701.2, C$_{1}$= 25.2~  and~  C$_{2}$=0.	\\
The photoelectric cross section at 10 keV is 
\begin{equation}
\sigma_{ph}(10 \textrm{keV})= 0.9532\cdot 10^{-24} \textrm{cm}^{-2}.
\end{equation}
At 10 keV the  E$^{-3}$ dependence is dominant, we thus extrapolate B.1. to higher energies using
\begin{equation}
\sigma_{ph}(\textrm{Energy})= \sigma_{ph}(10\, \textrm{keV})\times \Big(\frac{\textrm{Energy}}{10
\textrm{keV}}\Big)^{-3}.
\end{equation}
  We present in the upper panel of the Figure
  B.1. the value of the total absorption cross section as a function of energy.
The total absorption of  the emission from the central engine is characterized as 
\begin{equation}
Flux_{observed}= Flux_{source} \times e^{-\sigma_{Th}N_{H}} \times
e^{-\sigma_{ph}(E) N_{H}}.
\end{equation}
We compute this at 15 keV, the lower energy used in our study, and show the result in the lower panel of the
 Fig. B.1. An object is considered as strongly  absorbed when its flux has decreased by 50\% due to absorption effects.
 We see in the lower panel of Fig. B.1.
that this value corresponds to a column density around 70$\cdot$10$^{22}$\,cm$^{-2}$.\\
Therefore, we consider that below a column density of 70$\cdot$10$^{22}$cm$^{-2}$ the absorption is 
not an important effect at the energies covered in this study. Thus, we excluded objects with higher 
hydrogen column density.
\begin{figure}[!hbt]
\begin{center}
\includegraphics[width=11cm]{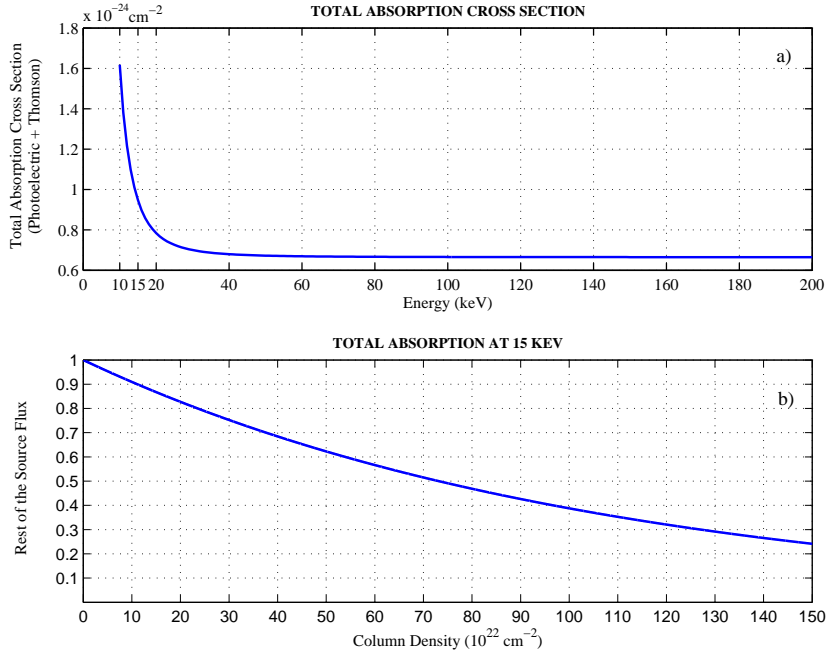}
\caption{Value and influence of the absorption at high energies. Fig. a gives the value
of the total absorption cross section composed of the photoelectric and Thomson components. Fig. b gives
the fraction of the rest flux  at 15 keV as a function of the column density.}
\end{center}
\end{figure}

\section{Average Counts Spectra fitted by the different models}
We present below the counts spectrum of each class fitted by the different models applied in this study.
\twocolumn
\begin{figure}[!h]
\centering
\includegraphics[width=6cm, angle=-90]{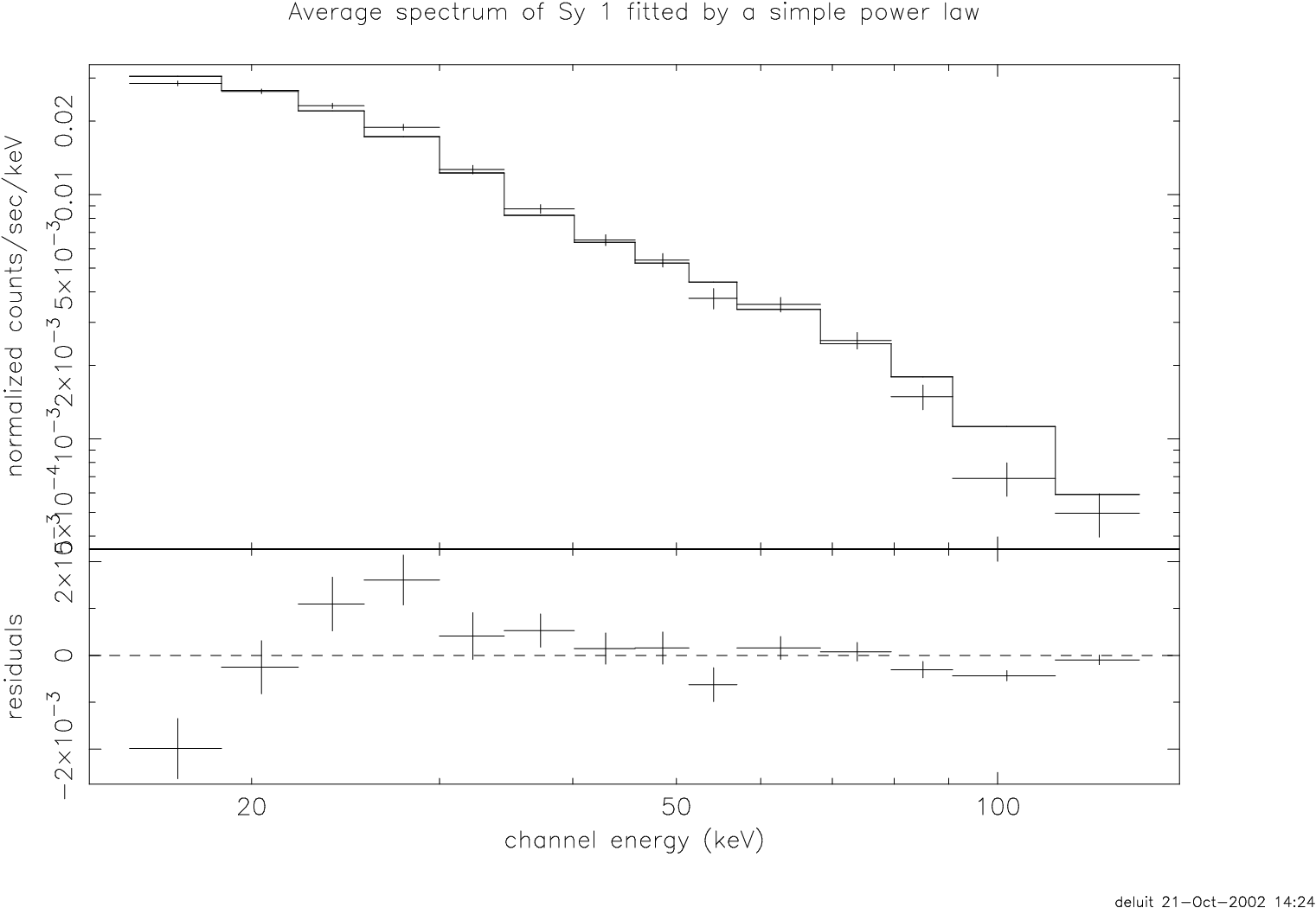}
\caption{Sy 1 average spectrum fitted by a power law model and  residuals }
\end{figure}

\begin{figure}[!h]
\centering
\includegraphics[width=6cm, angle=-90]{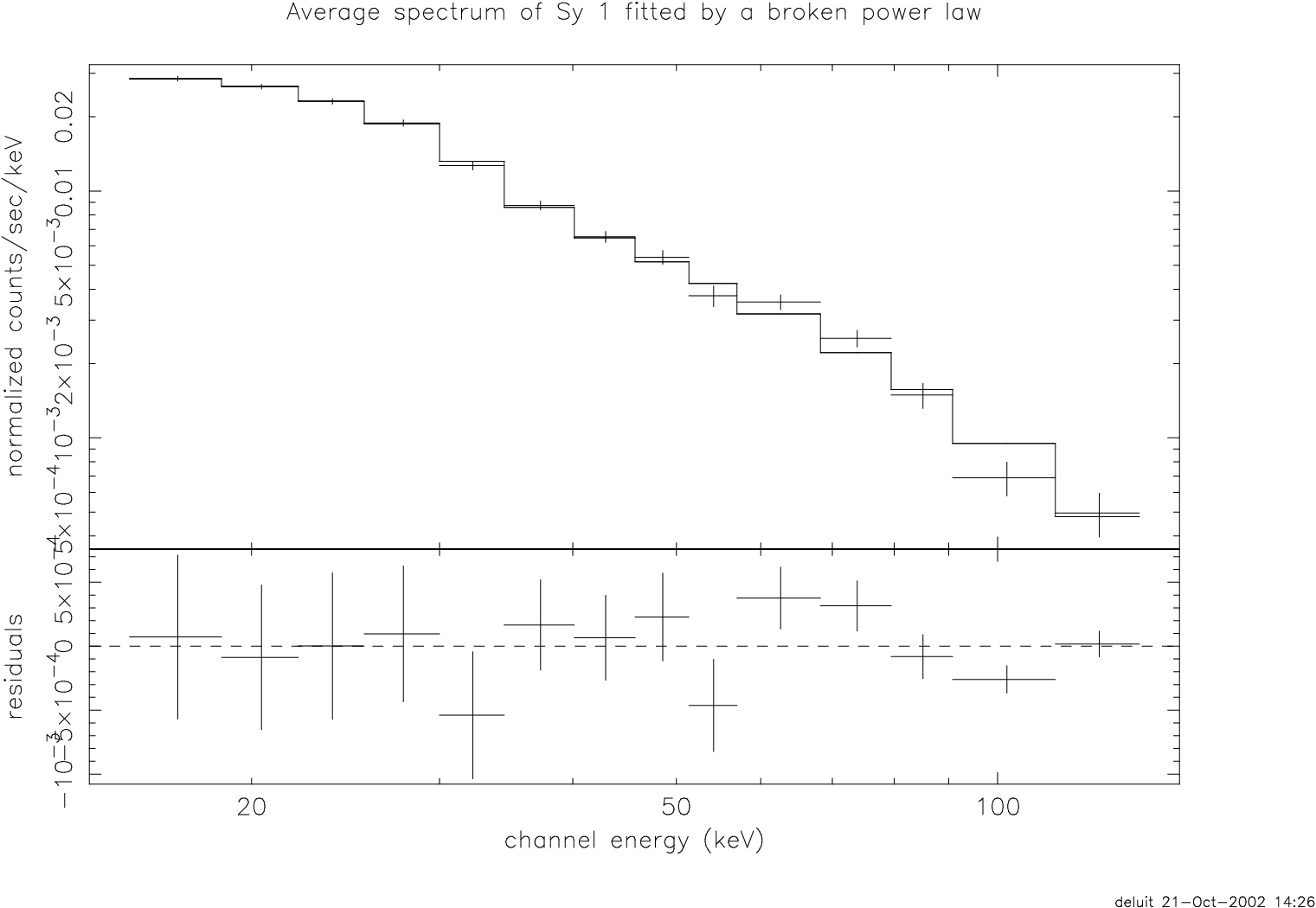}
\caption{Sy 1 average spectrum fitted by a broken power law model and  residuals }
\end{figure}

\begin{figure}[!h]
\centering
\includegraphics[width=6cm, angle=-90]{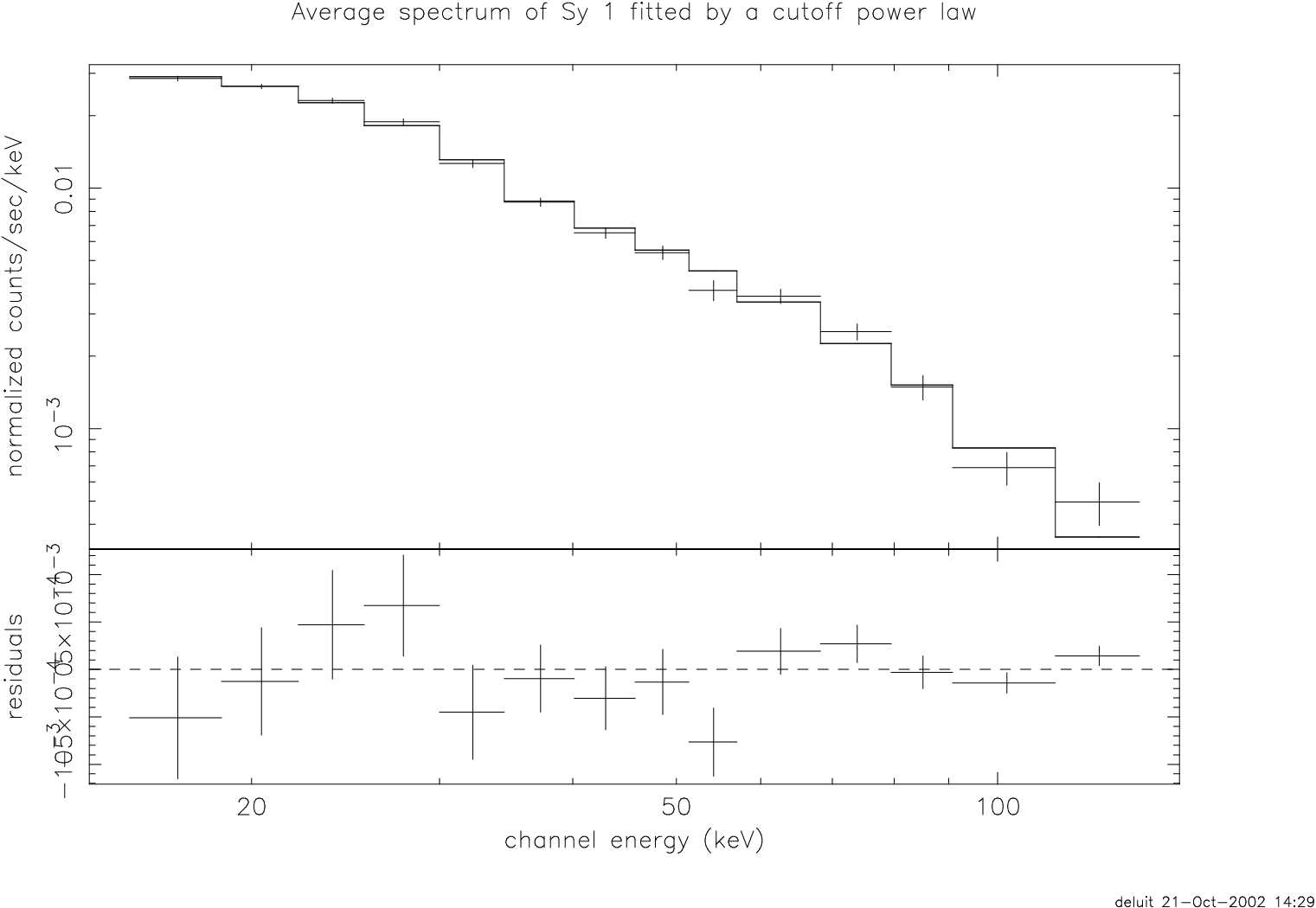}
\caption{Sy 1 average spectrum fitted by a cutoff power law model and  residuals  }
\end{figure}

\begin{figure}[!h]
\centering
\includegraphics[width=6cm, angle=-90]{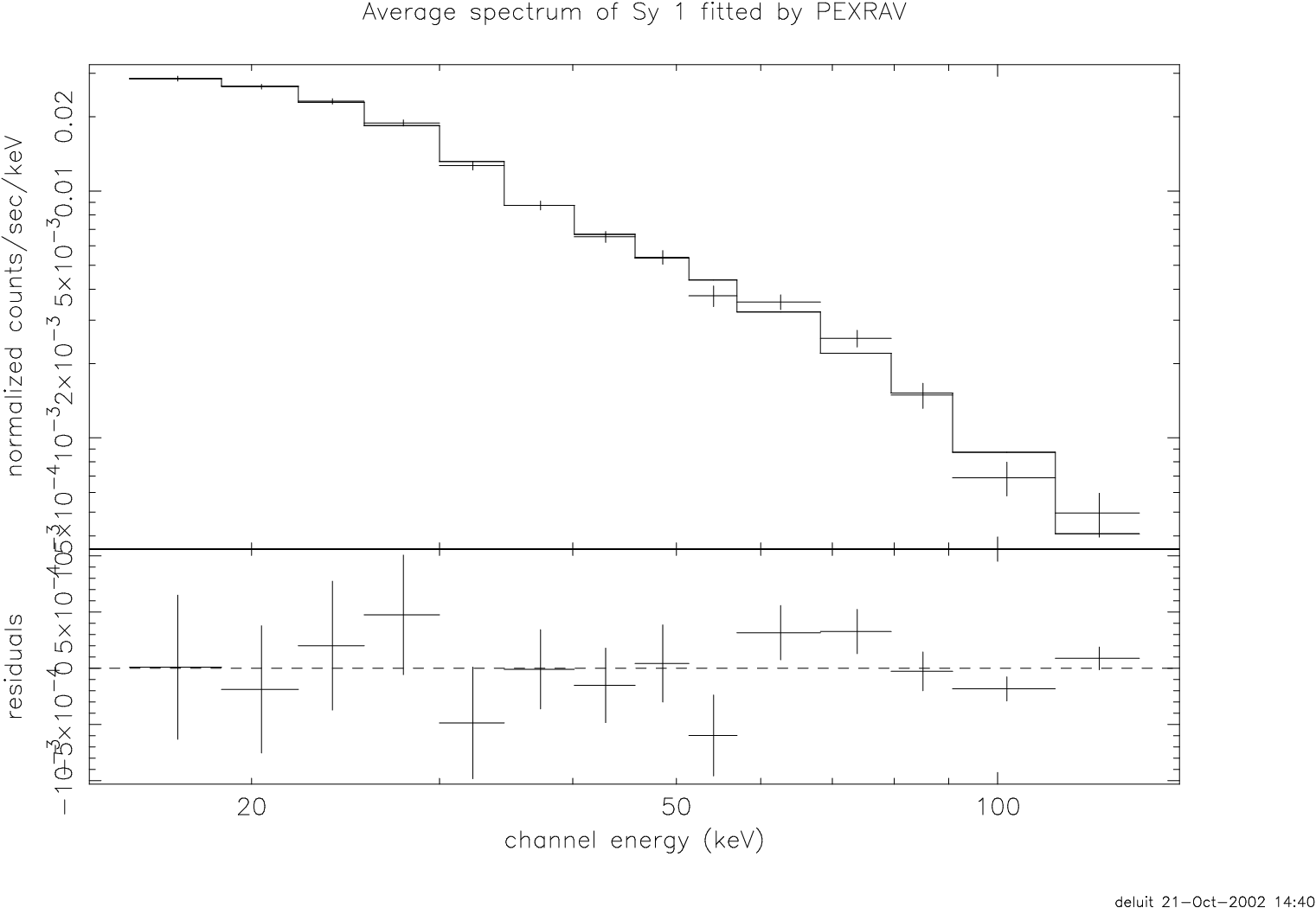}
\caption{Sy 1 average spectrum fitted by the PEXRAV model (cos $\theta$=0.45) and  residuals  }
\end{figure}

\begin{figure}[!h]
\centering
\includegraphics[width=6cm, angle=-90]{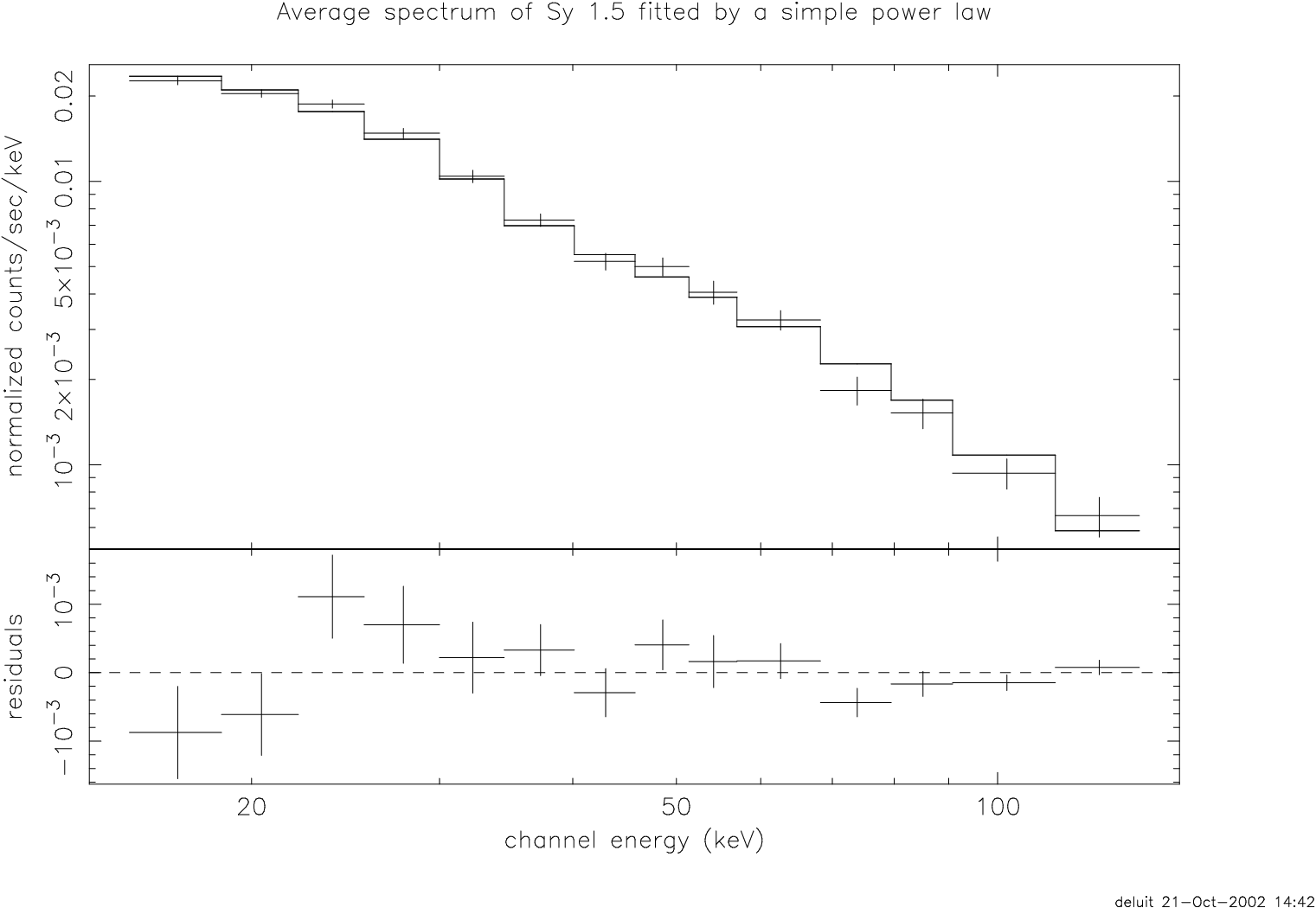}
\caption{Sy 1.5 average spectrum fitted by a power law model and residuals }
\end{figure}

\begin{figure}[!h]
\centering
\includegraphics[width=6cm, angle=-90]{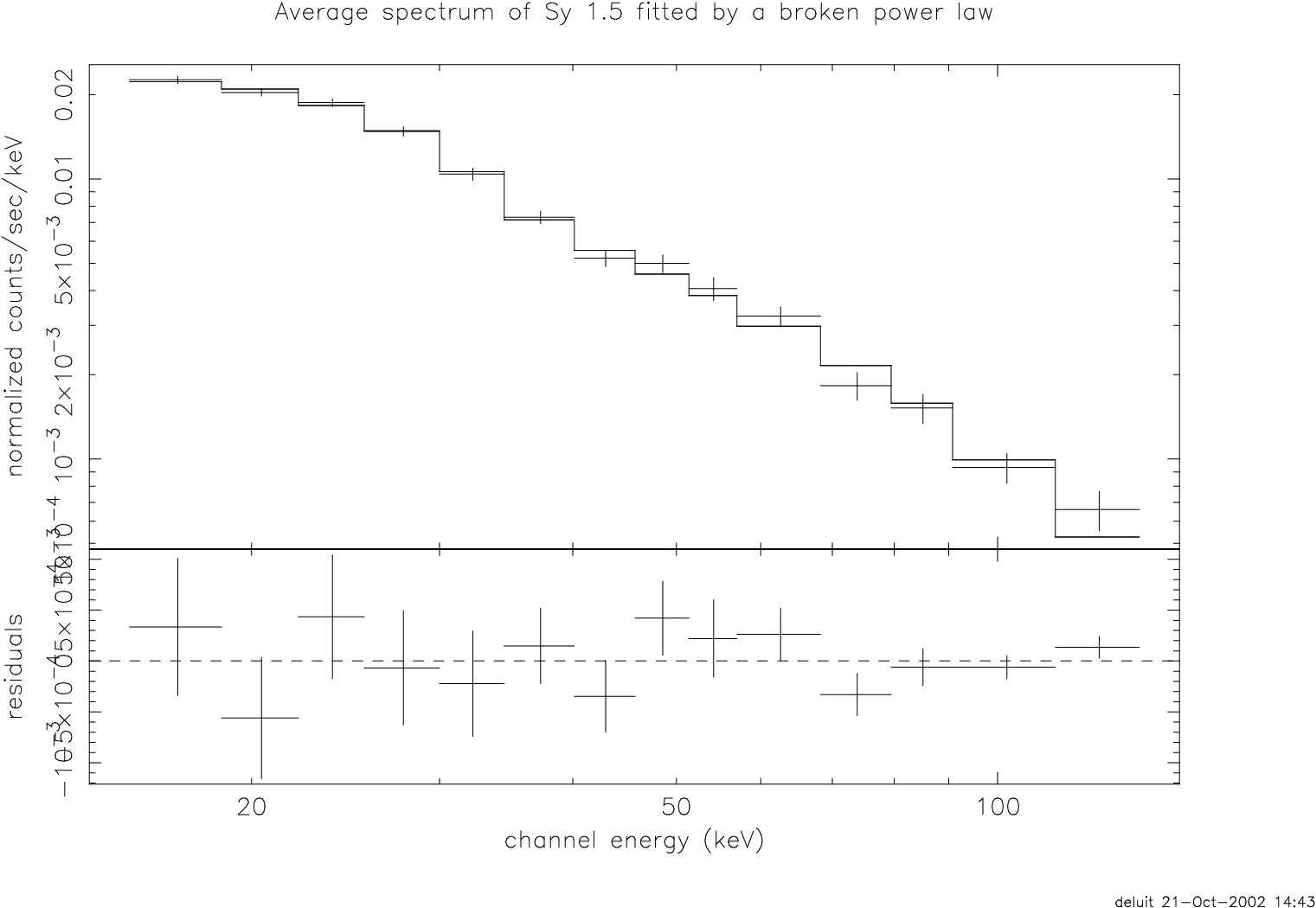}
\caption{Sy 1.5 average spectrum fitted by a broken power law model and  residuals }
\end{figure}

\begin{figure}[!h]
\centering
\includegraphics[width=6cm, angle=-90]{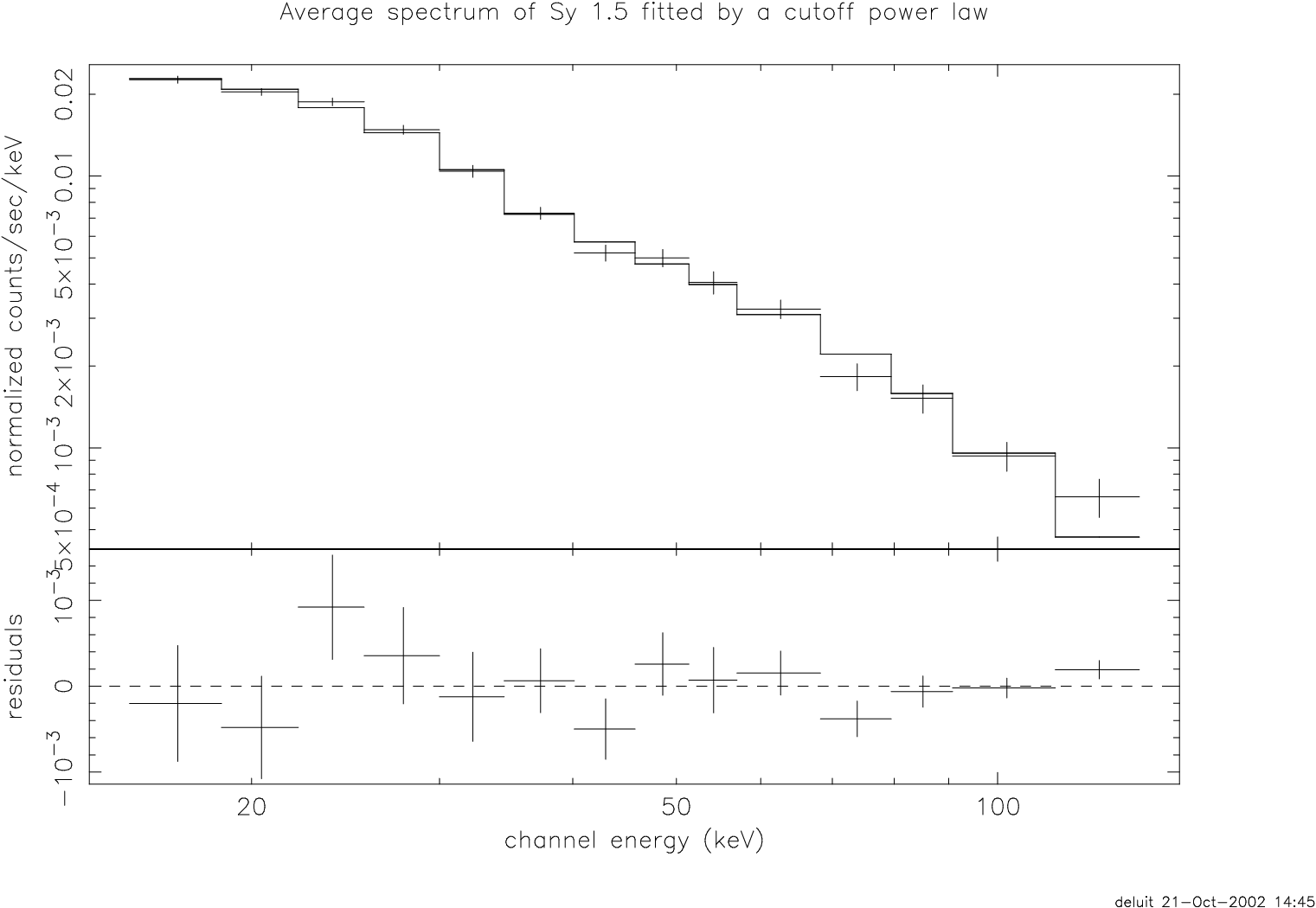}
\caption{Sy 1.5 average spectrum fitted by a cutoff power law model and  residuals }
\end{figure}

\begin{figure}[!h]
\centering
\includegraphics[width=6cm, angle=-90]{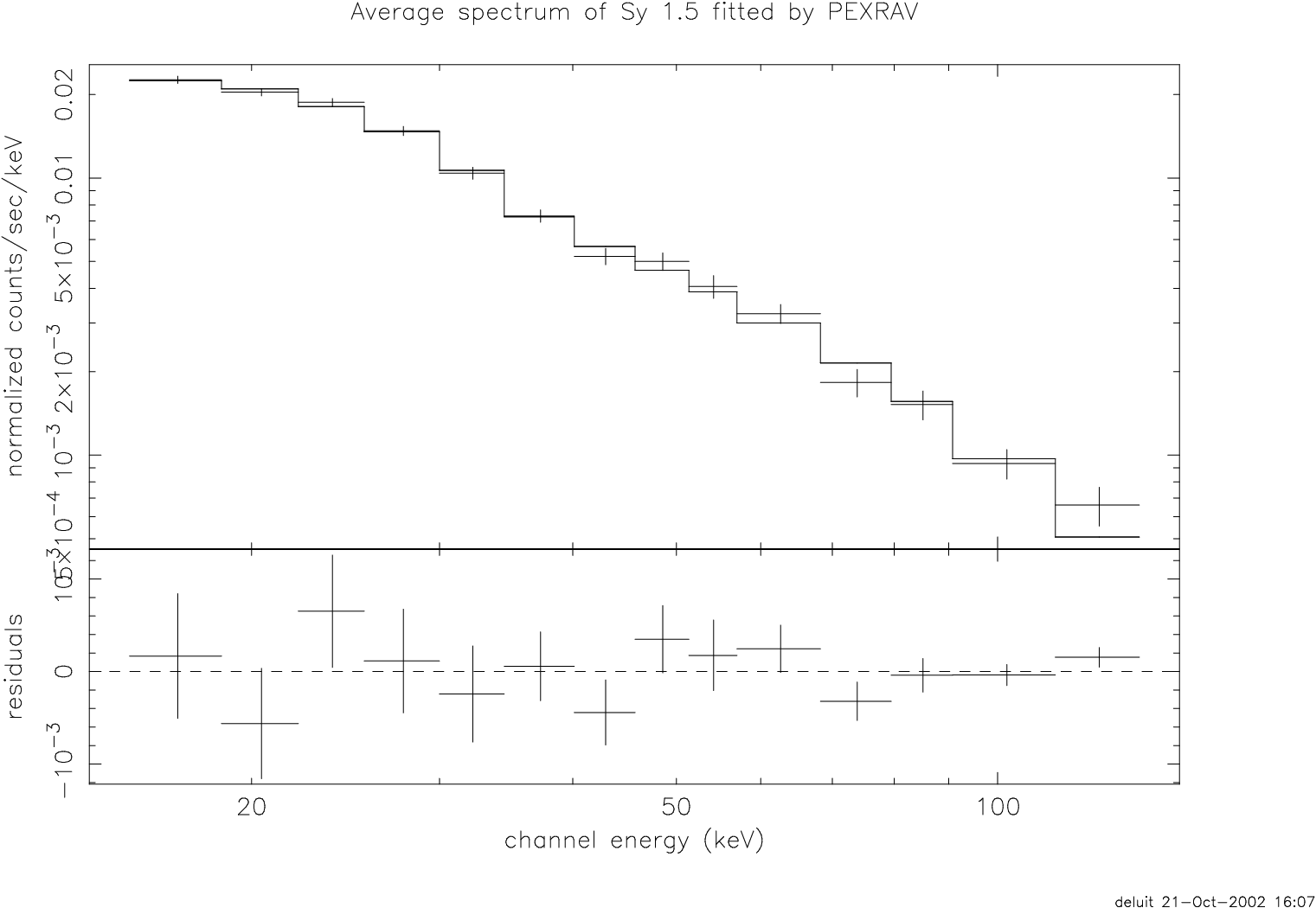}
\caption{Sy 1.5 average spectrum fitted by the PEXRAV model (cos $\theta$=0.45) and  residuals }
\end{figure}

\begin{figure}[!h]
\centering
\includegraphics[width=6cm, angle=-90]{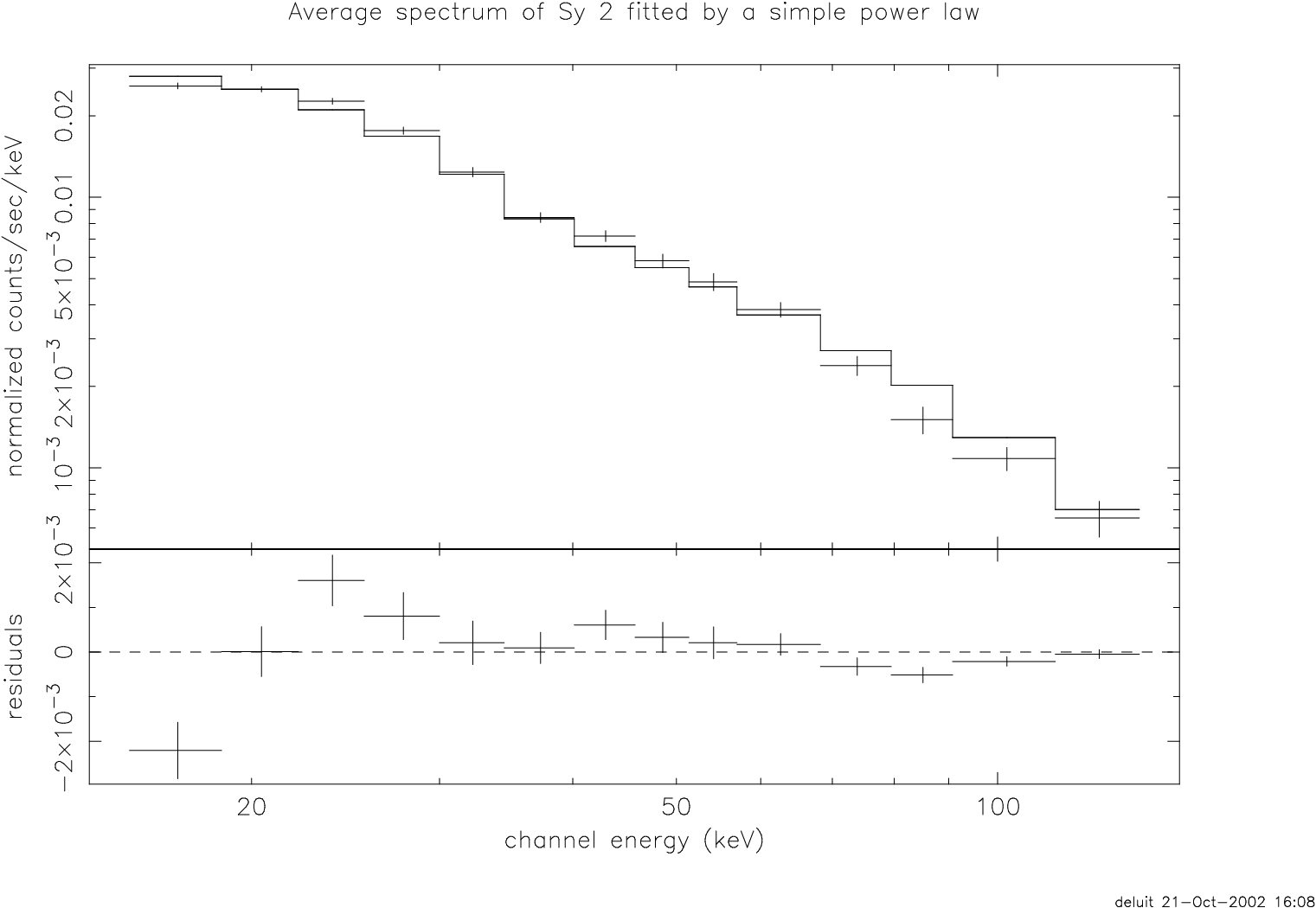}
\caption{Sy 2 average spectrum fitted by a  power law model and  residuals }
\end{figure}

\begin{figure}[!h]
\centering
\includegraphics[width=6cm, angle=-90]{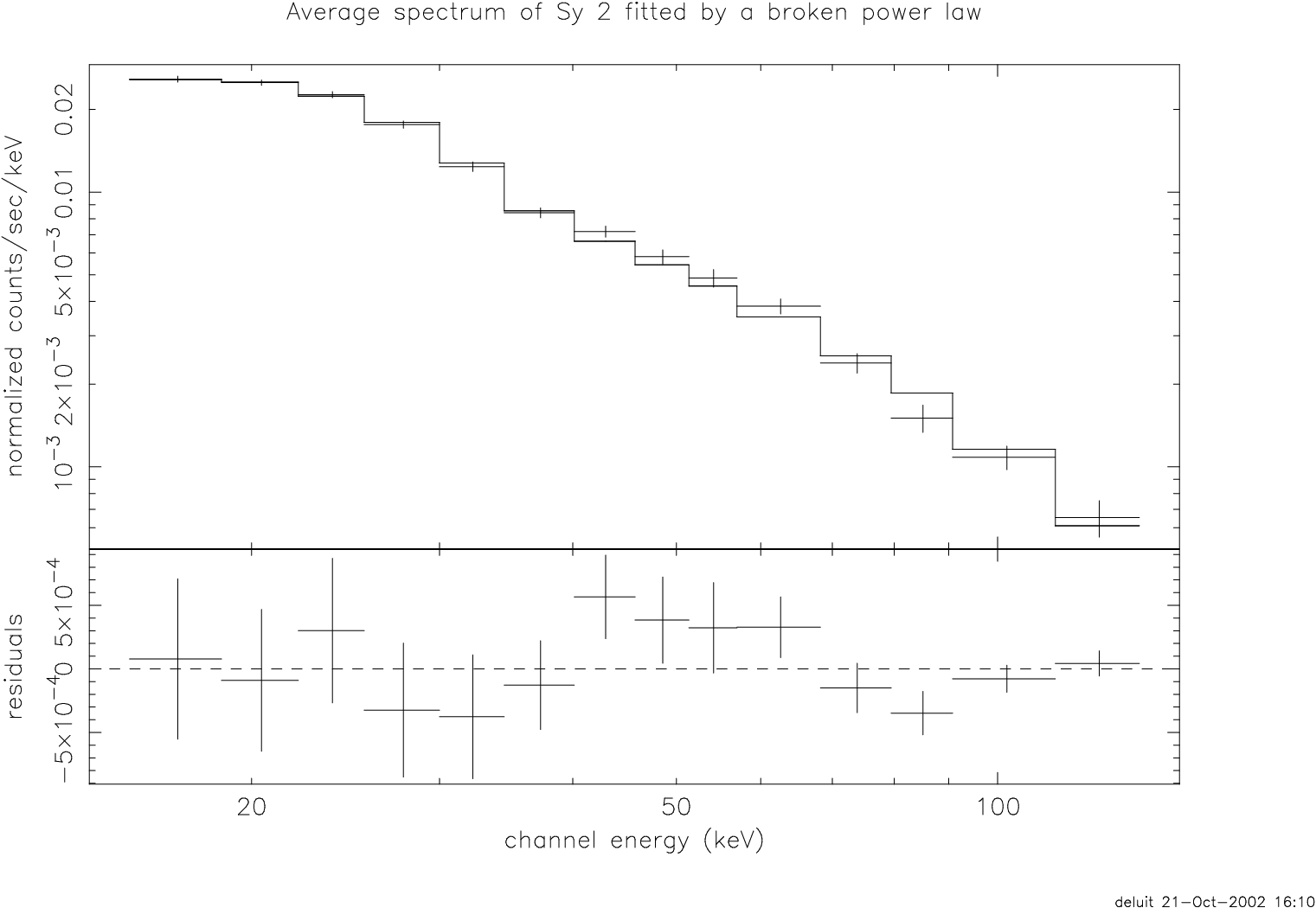}
\caption{Sy 2 average spectrum fitted by a broken  power law model and  residuals}
\end{figure}

\begin{figure}[!h]
\centering
\includegraphics[width=6cm, angle=-90]{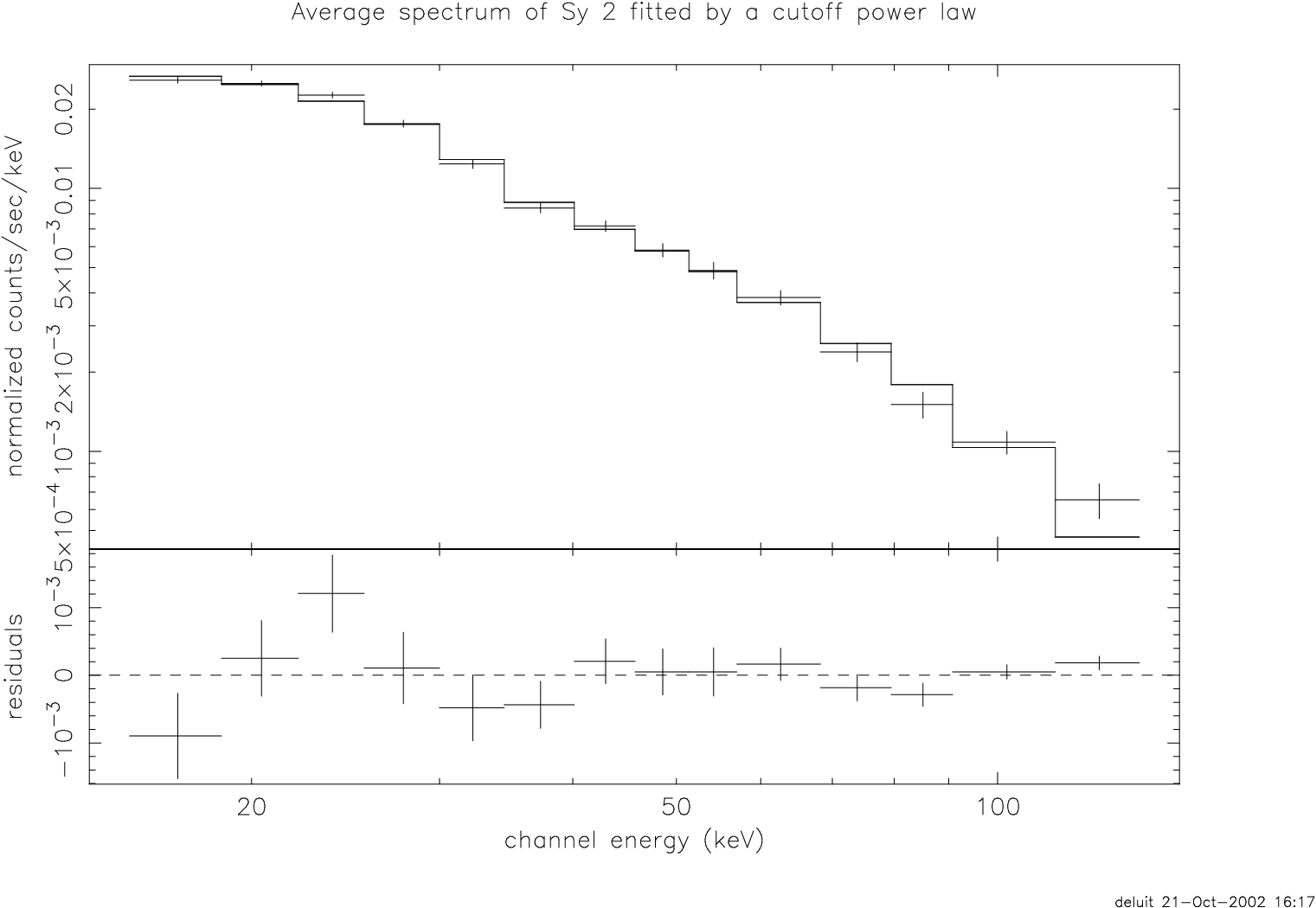}
\caption{Sy 2 average spectrum fitted by a cutoff power law model and  residuals }
\end{figure}

\begin{figure}[!h]
\centering
\includegraphics[width=6cm, angle=-90]{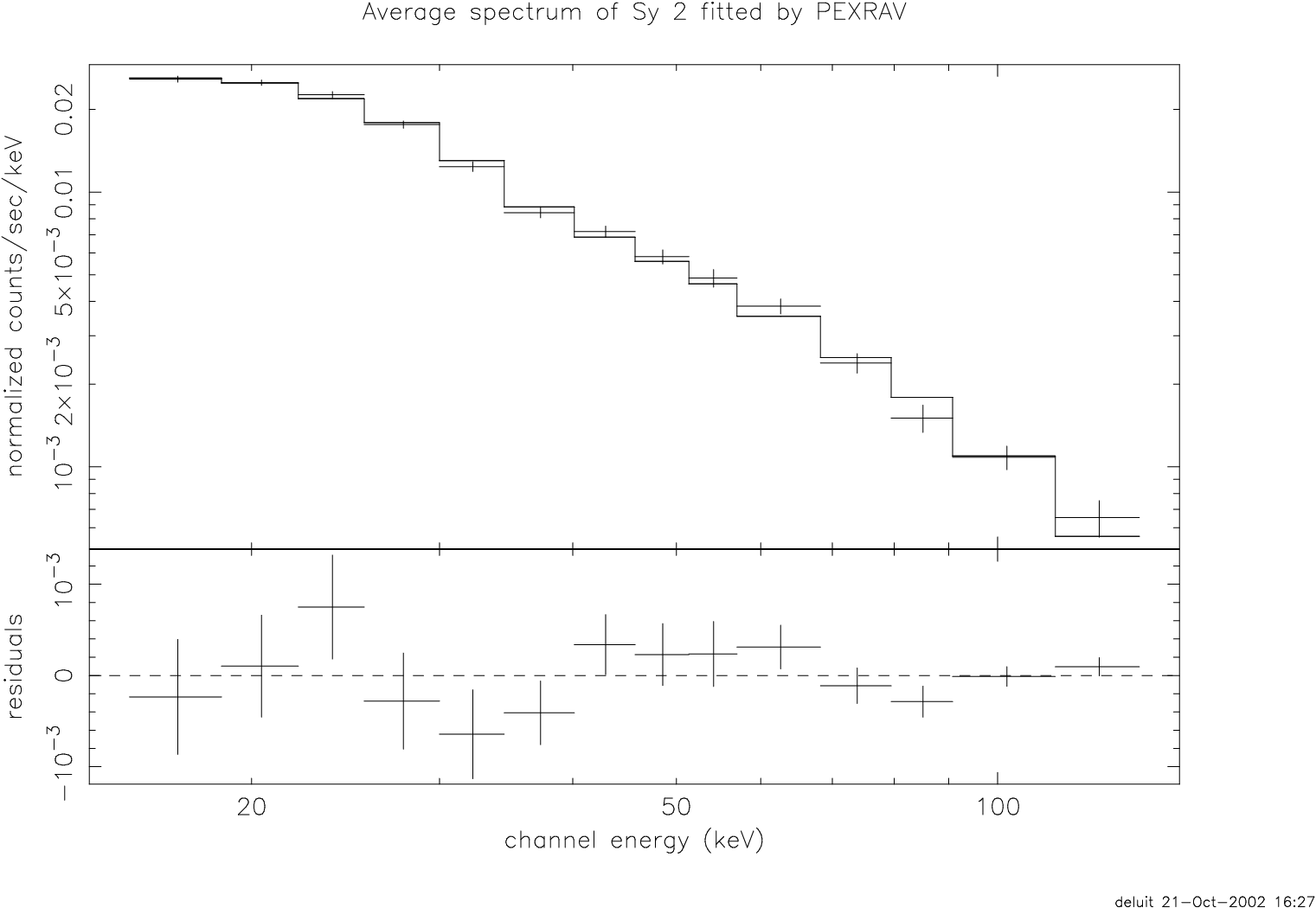}
\caption{Sy 2 average spectrum fitted by the PEXRAV  model (cos $\theta$=0.45) and  residuals}
\end{figure}
\end{document}